\documentclass[11pt]{article}
\usepackage{authblk}
\usepackage{amsmath}
\usepackage{amsfonts}
\usepackage{amssymb}
\usepackage{amstext}
\usepackage{tensor}
\usepackage{array}
\usepackage[margin=1in,footskip=0.25in]{geometry}
\usepackage[pdftex,
  pdftitle={Dark Energy or Modified Gravity? An Effective Field Theory Approach},
  pdfauthor={Bloomfield et al.},
  bookmarks,bookmarksopen=false,
  pdfstartview={FitH},
  linktocpage=true
]{hyperref}

\DeclareRobustCommand{\SkipTocEntry}[4]{}
\numberwithin{equation}{section}
\newcommand{\Mp}{m_0^2}

\title{Dark Energy or Modified Gravity? \\ {\it An Effective Field Theory Approach}}
\author[1,2]{Jolyon Bloomfield\thanks{jkb84@cornell.edu}}
\affil[1]{Center for Radiophysics and Space Research, Cornell University, Ithaca, NY 14853}
\affil[2]{Laboratory for Elementary Particle Physics, Cornell University, Ithaca, NY 14853}

\author[1,2]{\'Eanna \'E. Flanagan\thanks{eef3@cornell.edu}}

\author[3]{Minjoon Park\thanks{minjoonp@physics.umass.edu}}
\affil[3]{Department of Physics, University of Massachusetts, Amherst, MA 01003}

\author[4]{Scott Watson\thanks{gswatson@syr.edu}}
\affil[4]{Department of Physics, Syracuse University, Syracuse, NY 13244}

\date{\today}

\begin{document}
\maketitle
\thispagestyle{empty}

\begin{abstract}
  We take an Effective Field Theory (EFT) approach to unifying existing proposals for the origin of cosmic acceleration and its connection to cosmological observations.  Building on earlier work where EFT methods were used with observations to constrain the background evolution, we extend this program to the level of the EFT of the cosmological perturbations -- following the example from the EFT of Inflation. Within this framework, we construct the general theory around an assumed background which will typically be chosen to mimic $\Lambda$CDM, and identify the parameters of interest for constraining dark energy and modified gravity models with observations. We discuss the similarities to the EFT of Inflation, but we also identify a number of subtleties including the relationship between the scalar perturbations and the Goldstone boson of the spontaneously broken time translations.  We present formulae that relate the parameters of the fundamental Lagrangian to the speed of sound, anisotropic shear stress, effective Newtonian constant, and Caldwell's $\varpi$ parameter, emphasizing the connection to observations. It is anticipated that this framework will be of use in constraining individual models, as well as for placing model-independent constraints on dark energy and modified gravity model building.
\end{abstract}

\newpage

% Make a table of contents
\tableofcontents
\clearpage

%%%%%%%%%%%%%%%%%%%%%%%%%%%%%%%%%%%%%%%%%%%%%%%%%%%
%%%%%%%%%%%%%%%%%%%%%%%%%%%%%%%%%%%%%%%%%%%%%%%%%%%
%%%%%%%%%%%%%%%%%%%%%%%%%%%%%%%%%%%%%%%%%%%%%%%%%%%
\section{Introduction and Motivation}
As the accuracy of cosmological measurements continues to improve at a rapid pace, it is crucial to establish a robust and economic way to connect data with fundamental theory. In particle physics and condensed matter systems the framework of Effective Field Theory (EFT) has proven to be very successful at this endeavor (for reviews see \cite{Kaplan:2005es, Burgess:2007pt}). Recently, it has been argued that these methods can be useful in cosmology as applied to both inflation \cite{Weinberg:2008hq, Cheung2008, Senatore:2010wk} and late-time acceleration \cite{Creminelli2009, Park2010, Bloomfield:2011wa} (see \cite{Jimenez:2011nn, Jimenez:2012jg, Khosravi:2012qg} for related approaches).

In \cite{Weinberg:2008hq}, a covariant effective field theory for
single scalar field inflationary models was constructed, applicable
both to the inflationary background and to perturbations about that
background.
This approach was generalized to the case of late time acceleration in \cite{Park2010,Bloomfield:2011wa}.  A subtlety that arose in this generalization was the appearance of additional couplings and operators that involve matter and that give rise to modifications of gravity that can play an important role -- appearing as a scalar-tensor theory in the regime of observational interest.  Some observational constraints on
the parameters of this type of EFT were derived in \cite{Mueller2012}, who showed through an attractor analysis that a large class of possible models could be eliminated by the data.
Similar effective field theory analyses have also been applied to the
case of gravity coupled to a single vector field
\cite{ArmendarizPicon:2009ai}, as opposed to a scalar field.

This covariant, background-independent EFT approach has the goal
of being able to describe all possible background cosmologies as well as perturbations.  Because of this ambitious goal, it encompasses a smaller set of theories than an alternative EFT approach \cite{Cheung2008, Senatore:2010wk,Creminelli2009,Gubitosi2012} which fixes the background and attempts only to describe the theory of the perturbations.
In particular, theories where higher derivative operators give order unity corrections to the background cosmology fall outside its domain of validity \cite{Bloomfield:2011wa}.  For inflation this includes models such as DBI inflation \cite{Silverstein:2003hf}, and for dark energy model it includes models like K-essence \cite{ArmendarizPicon:2000dh} and ghost condensates \cite{ArkaniHamed:2003uy}.

The alternative approach that has been advocated in the context of single field \cite{Cheung2008} and multi-field \cite{Senatore:2010wk} inflation takes the background cosmology as given {\it a priori} and
constructs the EFT of perturbations around that background. For inflation, much like the standard model vacuum, this makes sense as the inflationary background is not directly observable. It was noted in \cite{Cheung2008} that an inflationary background spontaneously breaks the time translation invariance of the underlying theory. This often makes it possible to capture the observational predictions of a class of inflationary models through the study of a single scalar mode -- the Goldstone boson.  The advantage of this approach is not only the simplicity of a single degree of freedom, but the clear explanation for the existence of universal relations between many of the parameters of the low-energy theory -- such as those determining the sound speed and non-Gaussianity -- that arise because the Goldstone must non-linearly realize the time translation invariance.

In \cite{Creminelli2009}, the authors applied this approach to the study of quintessence models to address issues of stability when models approach $w<-1$. In this paper, we extend
the work of \cite{Creminelli2009} to allow more general couplings to matter, encompassing both dark energy and modified gravity models.
We do so by generalizing the derivation of the theory of \cite{Cheung2008} to allow for the presence of an arbitrary matter sector that is assumed to satisfy the weak equivalence principle.
We identify operators in the EFT that are interesting to study, and construct from those a general theory of scalar perturbations for dark energy and modified gravity. We also clarify some subtleties
that arise in the formalism when matter fields are present; in particular, the scalar mode associated with the dark energy field is no longer the Goldstone boson of the broken time translation symmetry.

With our theory in hand, we set about a program of
surveying the interesting physical effects associated with different operators in the EFT. We calculate the contributions of each operator to a set of observables consisting of the speed of sound, the anisotropic shear stress, the effective Newton's constant, Caldwell's $\varpi$ parameter
that parameterizes modifications to gravity \cite{caldwell2007},
and other phenomenological parameterizations of modified gravity used in the literature. We conclude by summarizing the observational effects of different operators, and present a list of which operators are included in different dark energy models.

Our work can be primarily viewed from two directions. From the effective field theory perspective, it provides a description of all operators that can be present, and a unified framework for investigating their effects. From the perspective of comparing dark energy models to observational data, it provides a unified framework for describing the perturbative behavior of different models. Each model can be matched to a certain set of operators, and with the framework developed here, the theory of linearized scalar perturbations within the context of the given model will be obtained. We anticipate that this formalism will be of use in constraining the size of different coefficients through numerical simulations, both in model-dependent and independent manners. Our presentation of observational quantities is constructed with this in mind, and we have included full details of the equations of motion in both synchronous and Newtonian gauges in order to facilitate numerical analyses.

While this paper was in preparation, Gubitosi \textit{et al.}
\cite{Gubitosi2012} published a very similar analysis. Their paper
introduces some of the ideas of this paper, and as such, we have taken
this into consideration when presenting our results. We refer readers
to their paper for further details that have been omitted from this
paper.  Some aspects of our work that do not overlap with Gubitosi
\textit{et al.} are our explanation of the modified role of the Goldstone
mode, the extent to which is valid to use a dark energy scalar field
as a ``clock'' when it is a subdominant component of the Universe at
high redshifts, the consideration of different operators in the EFT,
the inclusion of spatial curvature throughout, and the casting of final results in synchronous gauge. Where there is overlap, our analyses are in agreement.

The structure of the paper is as follows.
In Section \ref{sec:framework}, we present an overview of the effective field theory construction, starting from the unitary gauge. From the resulting action, we identify how the background evolution is specified, and describe operators that affect only perturbations. We transform out of unitary gauge in Section \ref{sec:pi}, introducing the scalar field that describes dark energy perturbations. We briefly relate the resulting general description to a vanilla quintessence model, before investigating the meaning of the scalar perturbation field. In Section \ref{sec:observations}, we undertake a program of extracting information relevant to observations from our general theory, presenting the calculation of a variety of quantities. We also link our results to phenomenological functions and dark energy models. The paper concludes with a summary, and directions for future investigations.

A number of detailed appendices accompany this paper. Appendix \ref{app:goldstone} details the application of the St\"uckelberg technique for
restoring full diffeomorphism invariance starting from a unitary gauge description.  Appendix \ref{app:Synchronous} gives some details of
gauge transformation computations.  We describe our conventions in momentum space in Appendix \ref{app:momentum}, and conclude by providing equations of motion in Appendix \ref{app:EOMs}.

%%%%%%%%%%%%%%%%%%%%%%%%%%%%%%%%%%%%%%%%%%%%%%%%%%%
%%%%%%%%%%%%%%%%%%%%%%%%%%%%%%%%%%%%%%%%%%%%%%%%%%%
%%%%%%%%%%%%%%%%%%%%%%%%%%%%%%%%%%%%%%%%%%%%%%%%%%%
\section{Summary of the
Effective Field Theory Framework for Cosmological Perturbations}
\label{sec:framework}

Our goal is to construct an EFT describing the perturbations about
a cosmological background solution of a theory containing gravity, a
single
scalar field, and a matter sector which is assumed to obey the weak
equivalence principle.  The theory should include all possible
couplings of the scalar to itself and to gravity, and will encompass
both modified gravity and dark energy models. Our approach is to follow the ideas of the EFT of Inflation \cite{Cheung2008} and the EFT of quintessence \cite{Creminelli2009}, generalizing to allow for arbitrary universal couplings to matter. In this section we describe the construction of the theory in unitary gauge.

In constructing the theory, we impose the weak equivalence principle,
so matter
fields only appear only in the matter action and all matter fields
couple to the same metric.
We emphasize that the weak equivalence principle is not a symmetry of the action, and thus
there is some tension between our assumption of this principle and our
use of the EFT theoretical framework.  In particular, if one assumes
the weak equivalence principle at the classical or tree level, it will
generically  be
violated by loop corrections \cite{PhysRevD.85.044052}.  Also, the
principle
excludes, for example, Yukawa couplings\footnote{Here we mean in the
  Jordan frame; in other frames such Yukawa couplings will be present.}
between the scalar field and fermionic
fields,
which would be natural from the EFT point of view because they
are consistent with the symmetries of the theory.
Nevertheless, we will proceed with the assumption of weak equivalence
principle, motivated
by the strong experimental evidence in its favor, at the $10^{-13}$
level for baryonic (non-dark) matter \footnote{We note however that for dark
matter the observational evidence for the weak equivalence principle
is much weaker, at the $\sim 10\%$ level \protect{\cite{Bean:2007ny}}.
This suggests an alternative theoretical approach to constructing
general dark energy models: to neglect
baryonic matter, assume that dark matter consists of a single massive
fermion, and construct the most general EFT
of that fermion coupled to gravity and a scalar field,
dropping the assumption of the weak equivalence principle.  We leave
investigation of this type of EFT for future work.}.

Given the assumption of the weak equivalence principle,
there exists a choice of conformal frame, the Jordan frame,
in which the matter couples to the metric $g_{\mu\nu}$ alone and not
to the scalar field.  We will find it convenient to work in this
frame, rather than in the Einstein frame, in part because it
simplifies computation of comparison with observational data.
With this choice of conformal frame, the
matter action is an arbitrary functional $S_m[g_{\mu\nu},\psi_m]$ of
the Jordan frame metric and the matter fields $\psi_m$.

We next specialize to a choice of time coordinate for which the
perturbation to the scalar field vanishes, that is, the class of
unitary gauges in which the scalar field perturbation is ``eaten'' by
the metric.  Such gauges are guaranteed to exist so long as the
background scalar field solution $\phi_0$ does not oscillate, i.e., as
long as its gradient is timelike and everywhere nonvanishing.  Models
in which the
background solution does oscillate, for example Refs.\ \cite{Ref3,Ref1,Ref2}, cannot be described by our
formalism.  This restriction is a limitation of our
background-dependent EFT
approach as compared to the covariant EFT approach of Weinberg
\cite{Weinberg:2008hq,Bloomfield:2011wa}.

This gauge specialization leaves unbroken a subgroup of the
diffeomorphism symmetry group consisting
of time-dependent spatial
diffeomorphisms of the form ${\bar t} = f(t)$, ${\bar x}^i = {\bar
  x}^i(t,x^j)$.
We now use the tools of effective field theory to construct the most
general action functional of a metric that is invariant under this
group of residual symmetries.
While the unitary gauge is not a useful gauge for investigating the
behavior of perturbations, the allowed objects are easily recognized,
and the action is remarkably straightforward in this gauge. We will transform later to more a more general class of gauges that makes
the scalar perturbation mode manifest.

The general unitary gauge action is given by
\begin{align}
\label{eq:theaction}
  S = \int d^4x \sqrt{-g} {}& \bigg[ \frac{\Mp}{2} \Omega(t) R + \Lambda(t) - c(t) \delta g^{00}
\nonumber \\
  & + \frac{M_2^4 (t)}{2} (\delta g^{00})^2
    + \frac{M_3^4 (t)}{3!} (\delta g^{00})^3
    + \ldots
\nonumber \\
  & - \frac{\bar{M}_1^3 (t)}{2} \delta g^{00} \delta \tensor{K}{^\mu_\mu}
    - \frac{\bar{M}_2^2 (t)}{2} (\delta \tensor{K}{^\mu_\mu})^2
    - \frac{\bar{M}_3^2 (t)}{2} \delta \tensor{K}{^\mu_\nu} \delta \tensor{K}{^\nu_\mu}
    + \ldots
\nonumber \\
  & + \lambda_1 (t) (\delta R)^2
    + \lambda_2 (t) \delta R_{\mu \nu} \delta R^{\mu \nu}
    + \gamma_1 (t) C_{\mu \nu \sigma \lambda} C^{\mu \nu \sigma \lambda}
    + \gamma_2 (t) \epsilon^{\mu \nu \sigma \lambda} \tensor{C}{_{\mu \nu}^{\rho \theta}} C_{\sigma \lambda \rho \theta}
    + \ldots
\nonumber \\
  & + m_1^2 (t) n^\mu n^\nu \partial_\mu g^{00} \partial_\nu g^{00}
    + m_2^2 (t) (g^{\mu \nu} + n^\mu n^\nu) \partial_\mu g^{00} \partial_\nu g^{00}
    + \ldots
    \bigg]
    + S_{m} [g_{\mu \nu}].
\end{align}
Here, the quantities $\delta g^{00}$, $\delta R$, $\delta R_{\mu \nu}$, and $\delta K_{\mu \nu}$ are perturbations (defined below) to the time-time component of the
metric, the Ricci scalar, the Ricci tensor, and the extrinsic
curvature tensor, respectively, and we use the metric signature $(-,+,+,+)$.  The quantity
 $C_{\mu \nu \sigma \lambda}$ is the Weyl tensor, and $n^\mu$ is the
 normal to surfaces of constant time. Each term has a coefficient that
 can be time-dependent, because the background solution breaks
time translation symmetry.
The mass $m_0$ multiplying the Ricci scalar is used to make $\Omega$ dimensionless, and is not necessarily the mass scale associated with the usual Planck mass.
The final line contains $S_m$, the matter action, which we leave arbitrary. The ellipsis indicates further terms that we have not included, at both quadratic and higher order in perturbations.\footnote{The coefficients are largely chosen to coincide with the literature. However, we do caution the reader that some of our coefficients are different from Refs. \cite{Cheung2008} and \cite{Gubitosi2012}. In particular, our $\Lambda(t)$ is defined differently (note that $c(t)$ is the coefficient of $\delta g^{00}$ instead of $g^{00}$). There also exist differing choices of sign convention for the normal and $K_{ij}$.}

In the action (\ref{eq:theaction}),
the perturbations to any quantity are defined by subtracting off that
quantity's value on the background written in terms of the metric,
normal, and functions of time. For example, we define
\begin{align}
  \delta K_{\mu \nu} = K_{\mu \nu} - K_{\mu\nu}^{(0)} = K_{\mu\nu} + 3H (g_{\mu \nu} + n_\mu n_\nu).
\end{align}
The ability to write the background components of operators in terms of dynamical variables is useful for maintaining symmetries in the action, and is unique to FRW backgrounds.

The action (\ref{eq:theaction}) depends on a spacetime metric, on the
background scalar field solution (equivalently a choice of time
coordinate $t$), and on some matter fields.  The equations of motion
are obtained by varying with respect to the metric and the matter
fields.  We note that the action (\ref{eq:theaction}) includes
only local spacetime operators, and so explicitly excludes theories
which contain nonlocal operators in this action\footnote{On the other
hand, nonlocal operators can arise starting from the local action
(\ref{eq:theaction}) when one integrates out the non-dynamical metric
variables (the lapse and shift) to obtain the analog of the Mukhanov-Sasaki
action for just the remaining scalar degree of freedom.  This occurs
in particular for standard single field inflation models with spatial
curvature \protect{\cite{Garriga:1999vw}}.  The EFT presented here
encompasses this type of non-locality.}.

The differences between the EFT of Inflation \cite{Cheung2008} action
and the action (\ref{eq:theaction}) are threefold\footnote{These three
  rules were independently derived by Gubitosi \textit{et al.}
  \cite{Gubitosi2012}}:

\begin{itemize}

\item Our action contains a matter action $S_m$.

\item Because we choose to describe the matter action in the Jordan frame, we now must include a function of time $\Omega(t)$ multiplying the Einstein-Hilbert term.  If we were to translate to an Einstein frame description, this function would parameterize the coupling of the scalar field to matter, as in a scalar tensor theory.

\item Terms such as $\delta R^2$ must now be included in the action.
In the EFT of inflation context, such terms could be eliminated
using field redefinitions, but those field redefinitions are no longer allowed since they would take us out of the Jordan frame.

\end{itemize}
\noindent
Note that all of these differences stem from the inclusion of matter.

The operators in the action split into two groups: those that contribute to the evolution of both the background and the perturbations, and those that only affect the perturbations. We will refer to the first group as the ``background'' terms (even though they also influence the perturbative behavior).  The background terms are given by the first line of the action \eqref{eq:theaction}, as well as the matter action.

\subsection{Background Evolution and Connection to \texorpdfstring{$\Lambda$}{Lambda-}CDM}

We begin by deriving how the background terms in the action (\ref{eq:theaction}), given by
\begin{align}
  S_0 = \int d^4x \sqrt{-g} {}& \bigg[ \frac{\Mp}{2} \Omega(t) R + \Lambda(t) - c(t) \delta g^{00} \bigg] + S_{m} [g_{\mu \nu}] \label{eq:minimalaction}\,,
\end{align}
relate to the background FRW solution, which is constrained by observations to be in close agreement with $\Lambda$CDM. Because the matter action couples universally to the metric $g_{\mu \nu}$ and no other interactions are present, the matter stress-energy tensor $T^{\mu\nu (m)}$ is conserved:
\begin{align}
  \nabla_\mu T^{\mu \nu (m)} = 0\,.
\end{align}
Here, the $(m)$ indicates a quantity associated with the matter fields.
At the level of the background, we can treat the matter as a perfect fluid described by an average energy density $\bar{\rho}_m$ and pressure $\bar{P}_m$, and so we have the usual continuity equation
\begin{align}
    \dot{\bar{\rho}}_m + 3H [\bar{\rho}_m + \bar{P}_m] = 0. \label{eq:mattercontinuity}
\end{align}
Here, and throughout the rest of the paper, overdots signify derivatives with respect to physical time\footnote{Because the remaining symmetry in unitary gauge permits redefinitions of the time coordinate, this formalism works equally well in conformal time.}.

A complete description of the background evolution entails knowledge of six functions of time: the scale factor $a(t)$,
the functions $\Omega(t)$, $\Lambda(t)$, and $c(t)$
that appear in the action (\ref{eq:minimalaction}), and the matter variables
$\bar{\rho}_m(t)$, and $\bar{P}_m (t)$. As we are constructing a theory of the perturbations, we require that the background evolution be specified \textit{a priori}, and so all of these functions need to be chosen. We can use the Friedmann equations to eliminate two of the six functions, for which we will select $\Lambda(t)$ and $c(t)$. Typically, the functions $a(t)$, $\bar{\rho}_m(t)$ and $\bar{P}_m(t)$ will be chosen to align closely with a $\Lambda$CDM evolution, but this does leave the function $\Omega(t)$ free \cite{Starobinsky:1998fr, Park2010}. This extra function of time describes, in the Jordan frame, a non-minimal coupling between the scalar field and the metric, and in the Einstein frame, a coupling of matter to the scalar field.

The Einstein equations for the background evolution are obtained by varying the action \eqref{eq:minimalaction} with respect to the metric. Using an FRW metric with spatial curvature $k_0$, the Friedmann equations are
\begin{align}
  3 \Mp \Omega \left[H^2 + \frac{k_0}{a^2} + H \frac{\dot{\Omega}}{\Omega} \right] &= \bar{\rho}_m - \Lambda + 2 c\,,
\\
  \Mp \Omega \left[3 H^2 + 2 \dot{H} + \frac{k_0}{a^2} + \frac{\ddot{\Omega}}{\Omega} + 2 H \frac{\dot{\Omega}}{\Omega} \right] &= - \Lambda - \bar{P}_m.
\end{align}
For a given $a(t)$, $\bar{\rho}_m(t)$, $\bar{P}_m(t)$ and $\Omega(t)$, we can obtain $c(t)$ and $\Lambda(t)$ from
\begin{align}
  c(t) &= - \frac{\bar{\rho}_m + \bar{P}_m}{2} - \Mp \Omega \left[ \dot{H} - \frac{k_0}{a^2} + \frac{1}{2} \frac{\ddot{\Omega}}{\Omega} - \frac{H}{2} \frac{\dot{\Omega}}{\Omega} \right]\,,
\\
  \Lambda(t) &= - \bar{P}_m - \Mp \Omega \left[3 H^2 + 2 \dot{H} + \frac{k_0}{a^2} + \frac{\ddot{\Omega}}{\Omega} + 2 H \frac{\dot{\Omega}}{\Omega} \right]\,.
\end{align}
We may also write the Friedmann equations in a more conventional form as
\begin{align}
  3 \Mp \Omega \left[H^2 + \frac{k_0}{a^2}\right] = \rho_Q + \bar{\rho}_m
, \qquad
  2 \Mp \Omega \left[\dot{H} - \frac{k_0}{a^2}\right] = - \left[\rho_Q + P_Q + \bar{P}_m + \bar{\rho}_m \right]\,, \label{eq:Friedmanntogether}
\end{align}
where $\rho_Q(t)$ and $P_Q(t)$ are the dark energy density and pressure respectively. The functions $c(t)$ and $\Lambda(t)$ can be related to these by
\begin{align}
  c(t) = \frac{\rho_Q + P_Q}{2} - \frac{\Mp \ddot{\Omega}}{2} + \frac{\Mp H \dot{\Omega}}{2}
, \qquad
  \Lambda(t) = P_Q - \Mp \ddot{\Omega} - 2 \Mp H \dot{\Omega}\,.
\end{align}
These equations describe how the background dark energy density and pressure evolve over the cosmological history. The terms proportional to $\dot{\Omega}$ and $\ddot{\Omega}$ are simply subtracting off the overcounting that comes from including the effective energy density and pressure from the non-minimal coupling $\Omega(t)$ in $\rho_Q$ and $P_Q$.

Combining the two Friedmann equations yields the background continuity equation.
\begin{align}
  \dot{\rho}_Q + \dot{\bar{\rho}}_m
  &=
  - 3 H \left[\rho_Q + P_Q + \bar{P}_m + \bar{\rho}_m \right] + 3 \Mp \dot{\Omega} \left[H^2 + \frac{k_0}{a^2}\right]
\end{align}
Subtracting off the background continuity equation for matter \eqref{eq:mattercontinuity} from this yields the continuity equation for dark energy.
\begin{align}
  \dot{\rho}_Q
  &=
  - 3 H \left[\rho_Q + P_Q\right] + 3 \Mp \dot{\Omega} \left[H^2 + \frac{k_0}{a^2}\right] \label{eq:darkcontinuity}
\end{align}

We now have an understanding of how the background evolution fixes the coefficients of the tadpole terms in the action. Before continuing to describe the remainder of the action, we pause to briefly discuss the importance of the background evolution. In previous work, a covariant EFT construction of the background evolution has been considered and perturbations of dark energy and matter have been examined \cite{Park2010, Bloomfield:2011wa}. It is well known that observations -- such as standard candle measurements to determine the expansion rate -- taken alone cannot constrain the background evolution, as a high level of degeneracy in the theoretical parameters remains.  The hope from the formulations \cite{Park2010, Bloomfield:2011wa} is thus to attempt to differentiate regions of parameter space that are degenerate at the level of background evolution by using a perturbative analysis.  However, by using a combination of theoretical and observational constraints some reduction in the allowed space of models is possible, as recently exemplified in \cite{Mueller2012}, where data in combination with an attractor analysis lead to some restrictions on models. Of note is that this analysis was unable to find a better fit to the observational data than the $\Lambda$CDM model. Thus, with the possible exception of early dark energy models that have a significantly modified cosmological evolution, using $\Lambda$CDM to approximately determine the behavior of the scale factor and matter energy density/pressure will provide a good description of the background.

This helps motivate the EFT approach to the perturbations that we are taking here, which by construction is agnostic to the origin of the background evolution. In the absence of a separate analysis of how the background evolution of a model depends on its parameters, we cannot link the perturbative behavior of the model to its background evolution. In a sense, we give up on the background evolution and focus on the perturbations. We will evaluate later the observational significance of scalar perturbations in dark energy models.

\subsection{Effective Theory of Perturbations in the Dark Sector}

Having described the background terms, we now briefly describe the remaining terms in the general action \eqref{eq:theaction}. Each of these terms consists of a time-dependent coefficient and products of objects that vanish on the FRW background.

The objects that we have available to us include $\delta g^{00}$, $\delta K_{\mu \nu}$, $\delta R_{\mu \nu \rho \sigma}$ (or $C_{\mu \nu \rho \sigma}$), $\delta R_{\mu \nu}$, and $\delta R$, where the perturbed quantities are defined above. Each operator is constructed from these objects pieced together with functions of time, $g^{\mu \nu}$, $n_\mu$, $\epsilon^{\mu \nu \sigma \lambda}$ and derivatives $\nabla_\mu$. By construction, the combination $n^\mu \delta K_{\mu \nu}$ is vanishing, and furthermore, the expression $R_{\mu \nu} n^\mu n^\nu$ can be rewritten using
\begin{align}
  \tensor{R}{_{\mu \nu}} n^\mu n^\nu = {}& K^2 - \tensor{K}{_\mu^\nu} \tensor{K}{_\nu^\mu} + \nabla_\mu (n^\nu \nabla_\nu n^\mu) - \nabla_\nu (n^\nu \nabla_\mu n^\mu).
\end{align}
Using these expressions and the ability to integrate by parts reduces the number of independent operators we need to consider.

A particular operator can be classified by its order in the perturbed objects. As we are interested only in linear perturbation theory, we will consider only operators that are quadratic in the perturbations. Even at quadratic order there are an infinite number of possible operators, because of the possibility of an arbitrary number of derivatives acting on the perturbed quantities. However, each derivative increases the scaling dimension of the operator, and so operators with more derivatives become less and less relevant.

It is convenient to split derivatives into space and time components using projection tensors. The projected space and time derivative operators are $(\delta^\mu_\nu + n^\mu n_\nu ) \nabla_\mu$ and $n^\mu n_\nu \nabla_\mu$ respectively. Operators involving time-projected derivatives typically lead to the appearance of higher order time derivatives in the equations of motion, indicating the presence of new fields (which are often ghosts from the Ostrogradski instability). Such terms are not problematic when treated within the context of an effective field theory \cite{Weinberg:2008hq}, so long as they do not promote non-propagating fields to propagating ones (for example, the vector fields in the metric).

Three-dimensional quantities such as the Riemann tensor of the spatial metric and contractions thereof do not need to be included in the EFT description, as such objects can be related to their four-dimensional counterparts through the Gauss-Codazzi relations \cite{Cheung2008}, which involve objects that have all been included in the EFT construction already. However, this does not preclude us from using the three-dimensional objects in our formalism, so long as we realize that they may not be independent from other operators. While from an EFT perspective, such terms are unnecessary, they can be useful for the purpose of matching to explicit theories of dark energy, and can also be useful for categorizing operators by separating out time derivatives from spatial derivatives in the four-dimensional gravitational invariants.

A number of examples of possible operators are given in Eq. \eqref{eq:theaction}. Given so many operators, each of which associated with its own time-dependent coefficient, it is of interest to determine which operators should be retained and which can be neglected. For this question, we are informed by three perspectives. First, what operators are required to describe linear perturbation theory for existing dark energy models using this formalism? Second, what operators give rise to unique signatures of interesting physics? Third, how do the observational effects of an operator scale with energy?

The third of these perspectives is of the most interest from the EFT point of view. Unfortunately, a complete analysis of this question is difficult to perform, as it is strongly dependent upon which operators dominate the kinetic dispersion relation of the scalar perturbations, and is also complicated by kinetic mixing with gravitational scalar perturbations. We intend to address this question in future work.

From our incomplete exploration of these questions, we propose that the following operators are most worth investigating (beyond the background operators):
\begin{align}
    & (\delta g^{00})^2\,,
&&
    \delta g^{00} \delta \tensor{K}{^\mu_\mu}\,,
&&
    (\delta \tensor{K}{^\mu_\mu})^2\,,
&&
    \delta \tensor{K}{^\mu_\nu} \delta \tensor{K}{^\nu_\mu}\,,
&&
    \delta g^{00} \delta R^{(3)}\,,
&&
    (g^{\mu \nu} + n^\mu n^\nu) \partial_\mu g^{00} \partial_\nu g^{00}
    \,.
    \label{eq:interestingops}
\end{align}
Here, $\delta R^{(3)} = R^{(3)} - 6 k_0 / a^2$, where $R^{(3)}$ is the Ricci scalar of the spatial metric $g_{ij}$ (including the scale factor). These operators are sufficient to describe linear perturbations about FRW cosmologies of very general single field dark energy models, including Horndeski's theory \cite{Horndeski1974, Deffayet2011, Kobayashi2011, Bloomfield2013}. In terms of physical signatures, they are sufficient to modify all cosmological signatures of GR that we have calculated. Finally, note that these operators are the leading order operators in terms of their mass dimension, assuming a na\"ive relativistic dispersion relation and ignoring kinetic mixing effects. The only operators not included at this order are $n^\mu n^\nu \partial_\mu g^{00} \partial_\nu g^{00}$ and $\delta g^{00} n^\mu \nabla_\mu \delta \tensor{K}{^\nu_\nu}$, which contribute higher order time derivatives to the equations of motion.

A number of these operators have been explored in the inflationary formalism, particularly with regards to correlation functions and stability. In particular, $(\delta g^{00})^2$, $(\delta \tensor{K}{^\mu_\mu})^2$ and $\delta \tensor{K}{^\mu_\nu} \delta \tensor{K}{^\nu_\mu}$ were investigated in Ref. \cite{Cheung2008}, while $(\delta g^{00})^2$ and $\delta g^{00} \delta \tensor{K}{^\mu_\mu}$ were analyzed in Ref. \cite{Creminelli2009}. The operators $(\delta g^{00})^2$ and $\delta g^{00} \delta \tensor{K}{^\mu_\mu}$ were investigated in Ref. \cite{Gubitosi2012}, who also suggested that $(g^{\mu \nu} + n^\mu n^\nu) \partial_\mu g^{00} \partial_\nu g^{00}$ may be of interest. The operator $\delta g^{00} \delta R^{(3)}$ has not been investigated before.

%%%%%%%%%%%%%%%%%%%%%%%%%%%%%%%%%%%%%%%%%%%%
\section{Effective Field Theory for Scalar Modes in Terms of St\"uckelberg Field}
\label{sec:pi}

In the previous section we showed that, in the unitary gauge, the EFT is specified by choosing a background solution (four free functions of time) and by choosing the time-dependent coefficients of a set of additional operators which only affect perturbations.  We would like now to examine
the predictions of this theory.  To this end, is convenient to use a different representation of the theory in which the diffeomorphism symmetry is restored and the dark energy field is explicit.

In this section we restore the full diffeomorphism symmetry in the action following Refs.\ \cite{Cheung2008,Creminelli2009}, by making use of the St\"uckelberg trick, which introduces a scalar field $\pi$, describing perturbations in the dark energy field. We express the action to second order in perturbations, and briefly demonstrate how this action can describe perturbations to vanilla quintessence models as an example. We then discuss some subtleties involved in the EFT construction in the present context where matter is present, and show that $\pi$ is not the Goldstone boson of broken time translations in this context.

We will specialize for the remainder of this paper to scalar
perturbation modes, neglecting vector and tensor modes.  This is
consistent at linear order in perturbations, where the three different
types of modes are uncoupled.  We will focus on linear perturbations
for most of this paper.
Going beyond linear order, it is
strictly speaking not consistent to include nonlinear effects in the
scalar degrees of freedom while neglecting the vector and tensor
degrees of freedom.  However since the vector and tensor modes
are expected to be small, it is likely that nonlinear couplings of the
scalar modes to these modes can be safely neglected.

\subsection{Restoring the Symmetry}\label{sec:stuckelberg}

Given the action in unitary gauge, we can restore time diffeomorphism invariance through the use of the St\"uckelberg technique, introducing a scalar field which nonlinearly realizes the symmetry. This procedure follows that given in the EFT of Inflation \cite{Cheung2008}, and we refer the reader to that paper for more details.

The basic premise is to perform the following coordinate transformation on the action,
\begin{align}
  \tilde{t} = t + \xi^0 (x^\mu)
, \qquad
  \tilde{x}^i = x^i\,,
\end{align}
where we adopt the convention that Greek letters represent four-dimensional indices, while Roman indices represent three-dimensional spatial indices, and the zero index is reserved for the time component. The spatial coordinates can be left unchanged, as the action still possesses spatial diffeomorphism invariance. Next, in the resulting action, replace  $\xi^0$ everywhere by $-\pi (\tilde{x}^\mu)$, effectively promoting $\xi^0$ to a dynamical field. The field $\pi$ transforms under time diffeomorphisms as $\pi \to \pi - \xi^0(x^\mu)$, so that $t + \pi$ is invariant. By construction, this will absorb any changes to the action caused by the change of coordinates. In effect, the action is now fully diffeomorphism invariant. The coordinates can then be relabeled to remove the tildes.

Using this procedure, explicit functions of time are modified according to
\begin{align}
f(t) \rightarrow f\left(t + \pi(x^\mu)\right) \,,
\end{align}
and will typically be Taylor-expanded in $\pi$. Operators that are not fully diffeomorphism invariant in four dimensions need to be transformed using the tensor transformation law. We describe this procedure in detail in Appendix \ref{app:goldstone}. After this procedure has been performed, the final action is a functional of a metric and a scalar $\pi$, and is invariant under time diffeomorphisms once again.

Upon introducing the $\pi$ field to the action \eqref{eq:theaction}, including the operators in \eqref{eq:interestingops}, we obtain the following action.
\begin{align}
  S = \int d^4x \sqrt{-g} {}& \bigg[ \frac{\Mp}{2} \Omega(t + \pi) R
  + \Lambda(t + \pi)
\nonumber \\
  & - c(t + \pi) \left(\delta g^{00} - 2 \dot{\pi} + 2 \dot{\pi} \delta g^{00} + 2 \tilde{\nabla}_i \pi g^{0i} - \dot{\pi}^2 + \frac{\tilde{g}^{ij}}{a^2} \tilde{\nabla}_i \pi \tilde{\nabla}_j \pi + \ldots \right)
\nonumber \\
  & + \frac{M_2^4 (t + \pi)}{2} \left(\delta g^{00} - 2 \dot{\pi} + \ldots \right)^2
\nonumber \\
  & - \frac{\bar{M}_1^3 (t + \pi)}{2} \left(\delta g^{00} - 2 \dot{\pi} + \ldots \right) \left(\delta \tensor{K}{^\mu_\mu} + 3 \dot{H} \pi + \frac{\tilde{\nabla}^2 \pi}{a^2} + \ldots \right)
\nonumber \\
  & - \frac{\bar{M}_2^2 (t + \pi)}{2} \left(\delta \tensor{K}{^\mu_\mu} + 3 \dot{H} \pi + \frac{\tilde{\nabla}^2 \pi}{a^2} + \ldots \right)^2
\nonumber \\
  & - \frac{\bar{M}_3^2 (t + \pi)}{2}
  \left(\delta \tensor{K}{^i_j} + \dot{H} \pi \delta^i_j
  + \frac{\tilde{g}^{i k}}{a^2} \tilde{\nabla}_k \tilde{\nabla}_j \pi + \ldots \right)
  \left(\delta \tensor{K}{^j_i} + \dot{H} \pi \delta^j_i
  + \frac{\tilde{g}^{j l}}{a^2} \tilde{\nabla}_l \tilde{\nabla}_i \pi + \ldots \right)
\nonumber \\
  & + \frac{\hat{M}^2 (t + \pi)}{2} \left(\delta g^{00} - 2 \dot{\pi} + \ldots \right) \left(\delta R^{(3)} + 4 H \frac{\tilde{\nabla}^2 \pi}{a^2} + 12 H \frac{k_0}{a^2} \pi + \ldots \right)
\nonumber \\
  & + m_2^2 (t + \pi) (g^{\mu \nu} + n^\mu n^\nu) \partial_\mu (g^{00} - 2 \dot{\pi} + \ldots) \partial_\nu (g^{00} - 2 \dot{\pi} + \ldots)
  + \ldots \bigg]
  + S_{m} [g_{\mu \nu}] \label{eq:basicpiaction}
\end{align}
The ellipsis in brackets reflect where we have truncated the expansions to quadratic order in perturbations. The final ellipsis represents other quadratic terms that we have neglected, as well as terms beginning at third order in perturbations. Tildes are used to denote quantities associated with the spatial metric. The time diffeomorphism symmetry is nonlinearly realized in this action. Gubitosi {\it et al.} \cite{Gubitosi2012} show how such an action can be converted into a covariant form.

\subsection{Metrics and Gauges}
Although Eq. \eqref{eq:basicpiaction} is written in terms of an arbitrary metric perturbation, it becomes convenient to choose a gauge associated with metric perturbations. Given the present formalism where time and space have been treated on a separate footing, we find it convenient to work in synchronous gauge, where $\delta g^{00} = g^{0i} = 0$. Well-known issues with this gauge will not arise for the purposes of our calculations. Furthermore, we envisage our equations being used in modified versions of software such as CAMB \cite{Lewis1999}, which uses synchronous gauge.

We use the following metric to describe perturbations in synchronous gauge
\begin{align}
  ds^2 = -dt^2 + a^2 (\tilde{g}_{ij} + h_{ij}) dx^i dx^j,
\end{align}
where $\tilde{g}_{ij}$ is the time-independent background spatial metric, and we let the spatial coordinate system be arbitrary.  The perturbation $h_{ij}$ is decomposed as
\begin{align}
  h_{ij} = \frac{h}{3} \tilde{g}_{ij} + \left( \tilde{\nabla}_i \tilde{\nabla}_j - \frac{\tilde{g}_{ij}}{3} \tilde{\nabla}^2 \right) \eta\,,
\end{align}
where tildes indicate quantities associated with the spatial metric. The metric determinant $\sqrt{-g}$ is
\begin{align}
  \sqrt{-g} = a^3 \sqrt{\tilde{g}} \left(1 + \frac{h}{2}\right).
\end{align}
Following Ma and Bertschinger \cite{ma1995}, we use an alternative definition of $\eta$ in momentum space, defined by
\begin{align}
  \eta(\vec{k}, t) &= - \frac{1}{k^2} [ h(\vec{k}, t) + 6 \tilde{\eta} (\vec{k}, t)].
\end{align}
We detail our conventions in momentum space in Appendix \ref{app:momentum}.

It will be convenient to discuss some phenomena in Newtonian gauge. In this gauge, we use the following metric.
\begin{align}
  ds^2 = -(1 + 2 \psi) dt^2 + a^2 (1 - 2 \phi) \tilde{g}_{ij} dx^i dx^j
\end{align}
We present the equations of motion for our theory in both gauges. For more information about our gauge choices, see Appendix \ref{app:Synchronous}.

\subsection{The action for the dark energy field}
We will later need the general-gauge action \eqref{eq:basicpiaction} in order to calculate the stress-energy tensor. But for the moment, we are interested in the dynamics of the $\pi$ field. Expanding \eqref{eq:basicpiaction} to second order in perturbations in synchronous gauge, the $\pi$ action becomes
\begin{align}
  S_\pi = \int d^4x a^3 \sqrt{\tilde{g}} {}& \bigg[
  c \left( \dot{\pi}^2 - \frac{\tilde{g}^{ij}}{a^2} \tilde{\nabla}_i \pi \tilde{\nabla}_j \pi \right)
  + \left( 3 \dot{H} c - \frac{\Mp}{4} \dot{\Omega} \dot{R}^{(0)} \right) \pi^2
  - c \dot{h} \pi
  + \frac{\Mp}{2} \dot{\Omega} \pi \delta R
\nonumber \\
  & + 2 M_2^4 \dot{\pi}^2
    + \bar{M}_1^3 \dot{\pi} \left(- \frac{\dot{h}}{2} + 3 \dot{H} \pi + \frac{\tilde{\nabla}^2 \pi}{a^2} \right)
    - \frac{\bar{M}_2^2}{2} \left(- \frac{\dot{h}}{2} + 3 \dot{H} \pi + \frac{\tilde{\nabla}^2 \pi}{a^2} \right)^2
\nonumber \\
  & - \frac{\bar{M}_3^2}{2}
  \left(- \frac{\tensor{\dot{h}}{^i_j}}{2} + \dot{H} \pi \delta^i_j
  + \frac{\tilde{g}^{i k}}{a^2} \tilde{\nabla}_k \tilde{\nabla}_j \pi \right)
  \left(- \frac{\tensor{\dot{h}}{^j_i}}{2} + \dot{H} \pi \delta^j_i
  + \frac{\tilde{g}^{j l}}{a^2} \tilde{\nabla}_l \tilde{\nabla}_i \pi \right)
\nonumber \\
  & - \hat{M}^2 \dot{\pi} \left(\delta R^{(3)} + 4 H \frac{\tilde{\nabla}^2 \pi}{a^2} + 12 H \frac{k_0}{a^2} \pi \right)
    + 4 m_2^2 \frac{\tilde{g}^{ij}}{a^2} \tilde{\nabla}_i \dot{\pi} \tilde{\nabla}_j \dot{\pi}
    + \ldots \bigg], \label{eq:piaction}
\end{align}
where we have used the background continuity equation \eqref{eq:darkcontinuity} in the form
\begin{align}
  \frac{\Mp}{2} \dot{\Omega} R^{(0)}
  + \dot{\Lambda}
  - 2 \dot{c}
  - 6 H c
  = 0\,, \label{eq:background}
\end{align}
to cancel tadpole terms and simplify the action. Here, $R^{(0)}$ represents the background Ricci scalar as a function of time, and $\delta R = R - R^{(0)}$, which is given by
\begin{align}
  \delta R &= 4 H \dot{h} + \ddot{h} - 2 \frac{k_0}{a^2} h + \frac{1}{a^2}\left(
  2 k_0 \tilde{\nabla}^2 \eta
  - \frac{2}{3} \tilde{\nabla}^2 h
  + \frac{2}{3} \tilde{\nabla}^2 \tilde{\nabla}^2 \eta
  \right)
\end{align}
in synchronous gauge. Note that there is a kinetic mixing between the gravitational perturbations and $\pi$, arising from a number of different operators. Because of this, we cannot read the speed of sound directly off this action. We present the equations of motion from this action in Appendix \ref{app:EOMs}.

\subsection{Example: The EFT of Quintessence}

In order to elucidate the $\pi$ field formalism, here we construct a simple quintessence model using our language. This example follows the work of Creminelli \textit{et al} \cite{Creminelli2009}.

Consider the following minimally coupled quintessence action.
\begin{align}
  S_\phi = \int d^4 x \sqrt{-g} \left[ \frac{m_P^2}{2} R - \frac{(\nabla \phi)^2}{2} - U(\phi) \right] \label{eq:quintessenceaction}
\end{align}
Assume some cosmological background solution $\phi = \phi_0(t)$ on an FRW metric, and ignore metric perturbations. The solution $\phi_0$ obeys the background equation of motion
\begin{align}
  \square \phi_0 - U^\prime (\phi_0) = 0. \label{eq:quintessencebackground}
\end{align}
The background pressure and energy density are given by
\begin{align}
  P_\phi = \frac{\dot{\phi}^2_0}{2} - U(\phi_0)
, \qquad
  \rho_\phi = \frac{\dot{\phi}^2_0}{2} + U(\phi_0).
\end{align}
Now, expand in perturbations as $\phi = \phi_0(t + \pi) = \phi_0(t) + \dot{\phi}_0 \pi + \frac{1}{2} \dot{\phi}_0^2 \pi^2 + \ldots$. In terms of the usual quintessence perturbation $\phi = \phi_0 + \delta \phi$, we see that $\delta \phi = \dot{\phi}_0 \pi$ to leading order in perturbations. Using this expansion in the action \eqref{eq:quintessenceaction}, we obtain
\begin{align}
  S_\phi = \int d^4 x \sqrt{-g} &\bigg[ \frac{m_P^2}{2} R
  - \frac{1}{2} g^{00} \dot{\phi}_0^2
  + (\square \phi_0) \left(\dot{\phi}_0 \pi + \frac{1}{2} \dot{\phi}_0^2 \pi^2\right)
  - \frac{1}{2} [\nabla (\dot{\phi}_0 \pi)]^2
\nonumber\\&
  - U(\phi_0)
  - U^\prime(\phi_0) \left(\dot{\phi}_0 \pi + \frac{1}{2} \dot{\phi}_0^2 \pi^2\right)
  - \frac{1}{2} U^{\prime \prime} (\phi_0) \dot{\phi}_0^2 \pi^2
  \bigg]\,,
\end{align}
after performing an integration by parts. The background equation of motion can be used to eliminate the terms in the large round brackets. After expanding derivatives, integrating by parts, using the time derivative of Eq. \eqref{eq:quintessencebackground} to simplify terms, writing $g^{00} = -1 + \delta g^{00}$ and using the background expressions for energy density and pressure, the resulting action is
\begin{align}
  S_\phi = \int d^4 x \sqrt{-g} &\bigg[ \frac{m_P^2}{2} R + P_\phi
  - \frac{\rho_\phi + P_\phi}{2} \delta g^{00}
  - \frac{\rho_\phi + P_\phi}{2} (\nabla \pi)^2
  + \frac{\rho_\phi + P_\phi}{2} 3 \dot{H} \pi^2
  \bigg].
\end{align}
By comparing the background terms to Eq. \eqref{eq:minimalaction}, we conclude that
\begin{align}
  \Lambda(t) = P_\phi = \frac{\dot{\phi}_0^2}{2} - U(\phi), \qquad
  c(t) = \frac{\rho_\phi + P_\phi}{2} = \frac{\dot{\phi}_0^2}{2}, \qquad
  \Omega(t) = 1\,,
\end{align}
for this model. The remaining terms then agree with our general expansion, Eq. \eqref{eq:piaction}, with all other operators identically vanishing.

\subsection{The EFT Construction in the Presence of Matter: Some Theoretical Considerations}
\label{sec:goldstone}

As described above, the basic ideas of the original construction of the EFT of Inflation \cite{Cheung2008} are as follows. First, specialize to the unitary gauge associated with the inflaton field, so that the perturbation to the inflaton field vanishes, and the only remaining field is the metric. Second, write down the most general action that respects the subgroup of the diffeomorphism symmetry group consistent with that gauge condition. Third, use the St\"uckelberg technique of introducing a new dynamical field, $\pi$, to restore full diffeomorphism invariance. The new dynamical field turns out to be the Goldstone boson associated with the breaking of time translation invariance by the background solution.

Consider now a more complicated context such as multifield inflation \cite{Senatore:2010wk}, or the case considered in this paper, where matter and a dark energy field are both present. In these cases, a number of questions about the EFT construction arise:
\begin{itemize}
\item Which unitary gauge do we use? For example, if we have two scalar fields $\phi$ and $\psi$, do we use the gauge for which $\delta \phi=0$, or the gauge where $\delta \psi=0$? If one field is the dominant component of the energy density in the Universe, must we use that field to define the gauge?
\item When we use the St\"uckelberg technique to restore full diffeomorphism invariance, is the scalar field $\pi$ that is used still the Goldstone boson associated with broken time translation invariance?
\end{itemize}

The answers to these questions are, first, that at a classical level it is legitimate to use unitary gauge associated with any field, even a subdominant one, to define the theory. The only complication is that the usual EFT rules for estimating the coefficients of operators in terms of a cutoff scale need not apply if one uses a subdominant field, as we discuss further below. Second, the field $\pi$ that arises from the St\"uckelberg technique is no longer the Goldstone mode in general.

To explain these issues, we temporarily specialize to a simple model system of a particle moving on a $N$-dimensional manifold whose action is
\begin{align}
S[x^A(\tau)] = \int d\tau \left\{ \frac{1}{2} g_{AB}[x^C(\tau)] \frac{d x^A}{d\tau} \frac{d x^B}{d\tau} - V[x^C(\tau)] \right\},
\end{align}
where $x^A = (x^1, \ldots, x^N)$. This system is not invariant under time reparameterizations. However we can rewrite it in a way that is by introducing the one dimensional metric $d\tau^2 = e^{2 \alpha(t)} dt^2$, in terms of which the action becomes
\begin{align}
S[x^A(t),\alpha(t)] = \int dt \left\{ \frac{1}{2} e^{-\alpha(t)} g_{AB}[x^C(t)] \frac{d x^A}{dt} \frac{d x^B}{dt} - e^{\alpha(t)} V[x^C(t)] \right\}.
\label{eq:particleaction}
\end{align}
This form of the action is explicitly invariant under the time reparameterizations
\begin{align}
t \to {\bar t} = {\bar t}(t),\ \ \ \ \
x^A(t) \to {\bar x}^A({\bar t}) = x^A(t({\bar t})),\ \ \ \ \
\alpha(t) \to {\bar \alpha}({\bar t}) = \alpha(t({\bar t})) + \ln
\left( \frac{dt}{d\bar t} \right).
\end{align}

Consider first the status of the Goldstone mode in this model system. Specialize for simplicity to the gauge
$\alpha=0$. Suppose that we have a background solution
\begin{align}
x^A(t) = x^{(0)\,A}(t)\,,
\end{align}
and we are interested in the theory of the perturbations $\delta x^A(t)$. The Goldstone mode $\pi$ associated with the broken time translation symmetry is a coordinate on field space defined by
\begin{align}
\delta x^A(t) = x^{(0)\,A}(t + \pi(t)) - x^{(0)\,A}(t)\,, \label{eq:goldstone}
\end{align}
for $1 \le A \le N$. To get a complete set of coordinates on field space one has to add $N-1$ other coordinates, say, $\xi^2, \ldots, \xi^N$. The Goldstone mode has the property that $\pi = $ constant, $\xi^A = 0$ for $2 \le A \le N$, is always a solution. The Goldstone mode is only defined when we are given a complete background solution.

Consider next the St\"uckelberg technique. This trick can be defined in very general contexts, when one does not specify a background solution, and so it cannot involve the Goldstone mode in general. As an example, let us fix the gauge in our model by demanding that
\begin{align}
x^1(t) = f(t)\,,
\end{align}
where $f(t)$ is some fixed function that we have picked (it could be a background solution for $x^1$). The action \eqref{eq:particleaction} becomes
\begin{align}
S[x^\Gamma(t),\alpha(t)] ={}& \int dt \left\{ \frac{1}{2}
 e^{-\alpha} g_{\Gamma\Sigma}[f(t),x^\Theta(t)] \frac{d
 x^\Gamma}{dt} \frac{d x^\Sigma}{dt}
+ e^{-\alpha} g_{\Gamma\,1}[f(t),x^\Theta(t)] \frac{d
 x^\Gamma}{dt} \frac{d f}{dt} \right\}
\nonumber \\
&
+ \int dt \left\{
\frac{1}{2} e^{-\alpha} g_{11}[f(t),x^\Theta(t)] \left(\frac{d f}{dt}\right)^2
- e^{\alpha(t)} V[f(t),x^\Theta(t)] \right\}.
\end{align}
Here Greek letters $\Gamma$ run over $2, \ldots, N$, so $x^A = (x^1,x^\Gamma)$. Note that now the field $\alpha$ is no longer pure gauge, it is physical.

We now want to restore time reparameterization invariance. Following the procedure outlined in \cite{Cheung2008}, we make a coordinate transformation $t \to {\tilde t} = t + \xi(t)$, rewrite all tensor components in terms of their transformed versions, replace $\xi$ everywhere by $\pi$, and then drop the tildes. This gives the action
\begin{align}
S[x^\Gamma(t),\alpha(t),\pi(t)] ={}& \int dt \left\{ \frac{1}{2}
 e^{-\alpha(t)} g_{\Gamma\Sigma} \frac{d
 x^\Gamma}{dt} \frac{d x^\Sigma}{dt}
+ e^{-\alpha(t)} g_{\Gamma\,1} \frac{d
 x^\Gamma}{dt} \frac{d}{dt} f(t + \pi(t))
\right\}
\nonumber \\
&
+ \int dt \left\{
\frac{1}{2} e^{-\alpha(t)} g_{11} \left(
\frac{d}{dt} f(t + \pi(t)) \right)^2
- e^{\alpha(t)} V \right\}\,,
\end{align}
where the dependence of the metric and potential on $[f(t+ \pi(t)), x^\Theta(t)]$ has been suppressed. By inspection we see that this is the same as the original action \eqref{eq:particleaction}, if we identify $f(t + \pi(t))$ with $x^1(t)$. Thus the St\"uckelberg trick does what we expect in this context: at a classical level, any unitary gauge can be used, and one can freely transform back and forth between the unitary gauge and general gauge representations of the theory.

Now suppose that one is given a background solution $x^{(0)\,A}(t)$. Then, for each $A$, one can define a unitary gauge by $x^{A}(t) = x^{(0)\,A}(t)$. Returning to a general gauge via the St\"uckelberg trick, for a given fixed $A$, will involve a field $\pi^A$ defined by
\begin{align}
x^A(t) = x^{(0)\,A}[t + \pi^A(t)].
\end{align}
The quantities $\pi^1, \ldots , \pi^N$ are a set of coordinates on the manifold that can be used instead of $x^1, \ldots, x^N$. In these coordinates the Goldstone mode is the diagonal excitation $\pi^1 = \pi^2 \ldots = \pi^N = \pi$.

Finally, consider a cosmological model with one dominant field and one subdominant field. If one uses the unitary gauge associated with the dominant field (i.e., uses that field as a clock), then the subdominant field can be completely neglected, and the situation reduces to the single field EFT case. Here, the standard EFT rules for the scaling of coefficients of operators with the cutoff of the EFT will apply. On the other hand, suppose one uses the unitary gauge associated with the subdominant field. At a purely classical level, this is a legitimate thing to do. However, because one is using a gauge that is poorly adapted to the dominant physics, the usual EFT scaling rules for coefficients of operators will no longer apply. Specifically, there is a new dimensionless quantity (roughly the ratio of energy densities in the dominant and subdominant fields\footnote{A more careful analysis suggests that the ratio of sum $P + \rho$ of each component may be the relevant quantity to consider. Note that this sum is responsible for breaking the time translation symmetry in the acceleration equation, \eqref{eq:Friedmanntogether}.}) on which the coefficients can depend. Therefore a subdominant field is not a very useful choice for a clock to define the gauge.

In the application of this paper, the energy densities of dark matter and dark energy are comparable today, and so our construction will be valid at low redshifts, with the usual EFT scaling rules being applicable. However, at high redshifts, the dark energy field which we are using as a clock will be a subdominant component.  This suggests that care must be taken when naively applying the EFT analysis to observations at high redshifts, since the breaking of time translation invariance at those redshifts is primarily due to dark matter and not dark energy. Understanding the EFT approach at high redshifts when dark energy is subdominant represents work in progress.

%%%%%%%%%%%%%%%%%%%%%%%%%%%%%%%%%%%%%%%%%
%%%%%%%%%%%%%%%%%%%%%%%%%%%%%%%%%%%%%%%%%
%%%%%%%%%%%%%%%%%%%%%%%%%%%%%%%%%%%%%%%%%
\section{Linking the Behavior of Perturbations to Observations} \label{sec:observations}

We now consider the full system of perturbations around the FRW background. We use the linearized equations of motion for scalar perturbations to calculate the speed of sound, Poisson equation, anisotropic stress, effective Newton's constant, and Caldwell's $\varpi$ parameter. We associate our constructions with phenomenological functions appearing in the literature, and conclude by highlighting the observational features of different models.

Some of these quantities have been calculated by Gubitosi \textit{et al.} \cite{Gubitosi2012}, using the operators $M_2$ and $\bar{M}_1$. Their calculations were mostly performed at the level of the action, by explicitly demixing the field content through field redefinitions. Our approach is related, but works entirely at the level of the equations of motion. The benefit to this is that we can compute physical quantities in terms of equations of motion with arbitrary coefficients. We then present the contribution that different operators make to these coefficients, allowing these quantities to be calculated for whichever combination of operators is desired.

\subsection{The Role of Perturbations}
As mentioned earlier, it is impossible to discriminate between different dark energy models based on their background evolution alone; we need to investigate how different models impact the evolution of perturbations, particularly the growth of large scale structure. Typically, this is influenced by the cosmological history, with little impact from dark energy perturbations. Once dark energy comes to dominate the universe (around $z \sim 2$), the perturbations may play a more prominent role, but are still unlikely to have a sufficiently noticeable impact on structure formation, with possible exceptions for small sound speeds. Unlike inflation, where the scalar perturbations set the initial conditions for future evolution, we do not expect to measure a correlation function associated with a dark energy scalar perturbation. It is thus worth asking why we are interested in the perturbations described by the $\pi$ field?

The answer is twofold. Firstly, understanding the dynamics of the perturbations is critical to controlling the stability of the theory. Even when dark energy is subdominant, an instability on a sufficiently small timescale can have catastrophic consequences. The issue of stability for quintessence has been investigated in some detail in Ref. \cite{Creminelli2009}.

Secondly, and more interestingly, when there exists a kinetic mixing between the scalar field and the gravitational scalar perturbations, the gravitational perturbations have their propagation modified. As $\pi$ perturbations in the form of $\ddot{\pi}$ source the gravitational scalar perturbations, the $\pi$ equation of motion, which contains second derivatives of the gravitational scalars, must be substituted in. This can cause significant modification to the dynamics of the gravitational perturbations. As these perturbations are also sourced by and react upon the matter perturbations, the behavior and growth of matter over- and under-densities are modified in this situation. Even when dark energy is a subdominant component of the universe, it can have a nontrivial effect on the growth of structure through this modification of the gravitational dynamics.

Therefore, this formalism is most powerful for constraining theories involving modified gravitational dynamics. In particular, we envisage that the size of some coefficients should be able to be constrained based on the drastic departure from general relativity that they entail. Conversely, for less ``exciting'' theories such as minimally coupled quintessence/$k$-essence, an analysis of large scale structure formation will provide information about the background evolution of the universe more so than the size of various coefficients. This is not particularly surprising, given the small number of operators necessary to describe minimally coupled quintessence and $k$-essence models. Nevertheless, the ability to constrain $\Lambda(t)$, $c(t)$ and $\Omega(t)$ is important, as these functions can describe basic deviations to the cosmological evolution as compared to $\Lambda$CDM.

This discussion highlights why the present formalism is an important extension of our previous work \cite{Park2010, Bloomfield:2011wa}, in which the regime of validity demanded that the operators corresponding to $\Omega$, $c$ and $\Lambda$ dominated the perturbative behavior, while other operators played a subdominant role.

\subsection{Speed of Sound}
We calculate the speed of sound of scalar perturbations in the sub-horizon limit by constructing the kinetic matrix for $\pi$, $h$ and $\tilde{\eta}$. The kinetic matrix is obtained by writing the $\pi$ equation of motion along with the time-time and space-space trace components of the Einstein equation as a matrix,
\begin{align}
  0 = \left(
  \begin{array}{ccc}
    \gamma_{\pi, \pi} & \gamma_{\pi, h} & \gamma_{\pi, \tilde{\eta}} \\
    \gamma_{tt, \pi} & \gamma_{tt, h} & \gamma_{tt, \tilde{\eta}} \\
    \gamma_{ss, \pi} & \gamma_{ss, h} & \gamma_{ss, \tilde{\eta}}
  \end{array}
  \right) \left(
  \begin{array}{c}
    \pi \\
    h \\
    \tilde{\eta}
  \end{array}
  \right). \label{eq:kineticmatrix}
\end{align}
In this matrix, $\gamma_{X,Y}$ is the coefficient of $Y$ in the $X$ equation of motion. Note that we use synchronous gauge for this calculation. Typically, this matrix will contain differential operators. To deal with these, replace $\partial_t \rightarrow - i \omega$.

The dispersion relationship is quite complicated in general. Here, we just focus on the geometric optics limit, in which terms with two derivatives are more important than terms with only one derivative, which in turn dominate over terms with no derivatives. At a given wavenumber, all coefficients will be expressed in terms of background quantities, and so should all be of the same order. For the purposes of calculating the speed of sound, we also neglect terms with three derivatives, although this assumption should be investigated carefully if detailed results are desired. The matrix resulting from this procedure is the kinetic matrix. Using the results given in Appendix \ref{app:EOMs}, we construct the contribution to the kinetic matrix for each operator we are considering, and present the matrices in Table \ref{tab:speedofsound}. The matrix for the background operators fixes the normalization and overall sign for each row (the normalization being otherwise arbitrary). In particular, we use Eqs. \eqref{eq:pimotion}, \eqref{eq:timetime} and \eqref{eq:SStrace} to extract the background contribution. The $\pi$ equation of motion has been left unchanged (coefficients can be read off from $\Delta_\pi$), while the gravitational equations of motion have been written as $0 = - \tensor{G}{^\mu_\nu} + \tensor{T}{^\mu_\nu}$.

\begin{table}[ht]
  \footnotesize
  \noindent\makebox[\textwidth]{
  \begin{tabular}{|>{$}c<{$}||>{$}c<{$}|}
    \hline
    \textbf{Operator} &
    \textbf{Kinetic Matrix Contribution}
    \\[1pt] \hline\hline
%
% Background terms
%
    \begin{aligned}
      \Omega(t), c(t), \Lambda(t)
    \end{aligned}
  &
    \begin{aligned}
      \Mp \left(
      \begin{array}{ccc}
        \frac{c}{\Mp} \left(\frac{k^2}{a^2} - \omega^2 \right) &
        \omega^2 \dot{\Omega} / 4 &
        \dot{\Omega} \frac{k^2}{a^2} \\
        \dot{\Omega} \frac{k^2}{a^2} &
        0 &
        - 2 \Omega \frac{k^2}{a^2} \\
        \dot{\Omega} \left(
      \frac{2}{3} \frac{k^2}{a^2}
      - \omega^2
      \right) &
        - \Omega \omega^2 /3 &
        - \frac{2}{3} \Omega \frac{k^2}{a^2}
      \end{array}
      \right)
    \end{aligned}
  \\[1pt] \hline
%
% (\delta g^{00)^2
%
    \begin{aligned}
      \frac{M_2^4 (t)}{2} (\delta g^{00})^2
    \end{aligned}
  &
    \begin{aligned}
      M_2^4 \left(
      \begin{array}{ccc}
        - 2 \omega^2 &
        0 &
        0 \\
        0 &
        0 &
        0 \\
        0 &
        0 &
        0
      \end{array}
      \right)
    \end{aligned}
  \\[1pt] \hline
%
% \delta g^{00} \delta \tensor{K}{^\mu_\mu}
%
    \begin{aligned}
      - \frac{\bar{M}_1^3 (t)}{2} \delta g^{00} \delta \tensor{K}{^\mu_\mu}
    \end{aligned}
  &
    \begin{aligned}
      \bar{M}_1^3 \left(
      \begin{array}{ccc}
        - \frac{1}{2} \left(H + 3 \frac{\dot{\bar{M}}_1}{\bar{M}_1}\right) \frac{k^2}{a^2} &
        \omega^2 / 4 &
        0 \\
        \frac{k^2}{a^2} &
        0 &
        0 \\
        - \omega^2 &
        0 &
        0
      \end{array}
      \right)
    \end{aligned}
  \\[1pt] \hline
%
% (\delta \tensor{K}{^\mu_\mu})^2
%
    \begin{aligned}
      - \frac{\bar{M}_2^2 (t)}{2} (\delta \tensor{K}{^\mu_\mu})^2
    \end{aligned}
  &
    \begin{aligned}
      \bar{M}_2^2 \left(
      \begin{array}{ccc}
        \frac{1}{2} \frac{k^4}{a^4} - 3 \dot{H} \frac{k^2}{a^2} & 0 & 0 \\
        3H \frac{k^2}{a^2} & 0 & 0 \\
        \left(H + 2 \frac{\dot{\bar{M}}_2}{\bar{M}_2} \right) \frac{k^2}{a^2} & - \omega^2 / 2 & 0
      \end{array}
      \right)
    \end{aligned}
  \\[1pt] \hline
%
% \delta \tensor{K}{^\mu_\nu} \delta \tensor{K}{^\nu_\mu}
%
    \begin{aligned}
      - \frac{\bar{M}_3^2 (t)}{2} \delta \tensor{K}{^\mu_\nu} \delta \tensor{K}{^\nu_\mu}
    \end{aligned}
  &
    \begin{aligned}
      \bar M_3^2 \left(
      \begin{array}{ccc}
       \frac{1}{2} \frac{k^4}{a^4} - \left(\dot H + \frac{k_0}{a^2}\right) \frac{k^2}{a^2} & 0 & 0 \\
       H \frac{k^2}{a^2}  & 0 & 0 \\
       \frac{1}{3} \frac{k^2}{a^2}\Big(H+\frac{2\dot{\bar M}_3}{\bar M_3} \Big) & - \omega^2 / 6 & 0
      \end{array}
      \right)
    \end{aligned}
  \\[1pt] \hline
%
% \delta g^{00} \delta R^{(3)}
%
    \begin{aligned}
      \frac{\hat{M}^2 (t)}{2} \delta g^{00} \delta R^{(3)}
    \end{aligned}
  &
    \begin{aligned}
      \hat{M}^2 \left(
      \begin{array}{ccc}
       2 \left( H^2 + \dot{H} + 2 H \frac{\dot{\hat{M}}}{\hat{M}}\right) \frac{k^2}{a^2} & 0 & 2 \left( H + 2 \frac{\dot{\hat{M}}}{\hat{M}}\right) \frac{k^2}{a^2} \\
       -4 \frac{k^2}{a^2} H & 0 & - 4 \frac{k^2}{a^2} \\
       0 & 0 & 0
      \end{array}
      \right)
    \end{aligned}
  \\[1pt] \hline
%
% g^{ij} \partial_i \pi \partial_h \pi
%
    \begin{aligned}
      m_2^2 (t) (g^{\mu \nu} + n^\mu n^\nu) \partial_\mu g^{00} \partial_\nu g^{00}
    \end{aligned}
  &
    \begin{aligned}
    \begin{aligned}
      m_2^2 \left(
      \begin{array}{ccc}
       - 4 \frac{k^2}{a^2} \omega^2 & 0 & 0 \\
       0 & 0 & 0 \\
       0 & 0 & 0
      \end{array}
      \right)
    \end{aligned}    \end{aligned}
  \\[1pt] \hline
  \end{tabular}}
  \normalsize
  \caption[Contributions to the Speed of Sound]{
    The contribution to the kinetic matrix from various operators.
  }
\label{tab:speedofsound}
\end{table}

To calculate the dispersion relationship, we take the determinant of the kinetic matrix, and set the result to zero. The solutions to the resulting equation give the dispersion relations for the propagating degrees of freedom.

As an example, the determinant of the kinetic matrix from the background operators alone yields
\begin{align}
  \frac{k^2}{a^2} \omega^2 \left[\left(c + \frac{3}{4} \Mp \frac{\dot{\Omega}^2}{\Omega}\right) \omega^2 - \left(c + \frac{3}{4} \Mp \frac{\dot{\Omega}^2}{\Omega}\right) \frac{k^2}{a^2}\right] = 0 \label{eq:firstdispersion}
\end{align}
with a meaningless multiplicative constant, indicating that there is only one propagating degree of freedom, with speed of sound unity. The presence of the operator $\tensor{M}{_2}(t)$ modifies the dispersion relationship as
\begin{align}
  c_s^2 = \frac{c + \frac{3}{4} \Mp \frac{\dot{\Omega}^2}{\Omega}}
           {c + \frac{3}{4} \Mp \frac{\dot{\Omega}^2}{\Omega} + 2 M_2^4} \,.
\end{align}
Including further operators can create particularly convoluted expressions for the speed of sound. It is of interest to note that the dispersion relationship arising from including the operators $\bar{M}_2$ and $\bar{M}_3$ is more complicated than a relativistic scaling due to the presence of fourth order spatial derivatives in the action, although when the coefficients are set to the negative of each other, the nonrelativistic terms cancel. The operator $m_2$ provides a rather unusual contribution to the dispersion relationship, which can no longer be interpreted as a typical speed of sound. A more careful analysis of dynamics in the presence of this operator is necessary if its effect on the scalar dynamics is desired.

It is possible, in principle, to apply a unitary transformation to the fields $\pi$, $h$ and $\tilde{\eta}$ to demix the fields at the level of the action, as described in Ref. \cite{Gubitosi2012}. The kinetic term in the action would then look like
\begin{align}
  S_{\mathrm{kinetic}} = \int d^4 x \sqrt{-g} \left[ f(t) \dot{\hat{\pi}}^2 - g(t) \frac{\tilde{g}^{ij}}{a^2} \tilde{\nabla}_i \hat{\pi} \tilde{\nabla}_j \hat{\pi} \right]\,, \label{eq:SoSaction}
\end{align}
where $\hat{\pi}$ is the propagating field. This form is of interest because it facilitates investigations of stability. The speed of sound can be read directly from this action as $c_s^2 = g(t)/f(t)$. However, there is an ambiguity in relating $f(t)$ and $g(t)$ to the coefficients of $\omega^2$ and $k^2/a^2$ in Eq. \eqref{eq:firstdispersion} and similar dispersion results, as $f \rightarrow f h(t), g \rightarrow g/h(t)$ will yield the same dispersion relationship. We can fix this ambiguity by taking the limit where $\Omega \rightarrow 1$ and only the background operators are turned on, so that we know that $f(t) = g(t) = c(t)$ from the action \eqref{eq:piaction}. When the dispersion relation changes, the kinetic action for the propagating degree of freedom can be constructed in a similar way, although this will require the addition of new operators to the action \eqref{eq:SoSaction}.

The speed of sound is intimately linked to questions of stability. Here, we comment only very briefly on some salient points. By requiring that the theory is ghost-free, the coefficient $f(t)$ of $\dot{\hat{\pi}}^2$ in Eq. \eqref{eq:SoSaction} must be positive. For just the background operators, this requires
\begin{align}
  c + \frac{3}{4} \Mp \frac{\dot{\Omega}^2}{\Omega}
  = \frac{\rho_Q + P_Q}{2} - \frac{\Mp \ddot{\Omega}}{2} + \frac{\Mp H \dot{\Omega}}{2} + \frac{3}{4} \Mp \frac{\dot{\Omega}^2}{\Omega} > 0\,,
\end{align}
which indicates that the equation of state of dark energy can become phantom-like ($w_Q < - 1$) without causing perturbations to become ghosts, for a suitable choice of $\Omega(t)$. The presence of further operators creates more mechanisms to avoid ghosts when the equation of state becomes phantom-like. If the numerator $g(t)$ becomes negative, then a gradient instability arises. Furthermore, a Jeans instability can also arise from certain operators. These issues were explored in detail in Ref. \cite{Creminelli2009}.

\subsection{Poisson Equation and Anisotropic Shear Stress}
The Poisson equation combines the time-time and time-space components of the gravitational equations of motion in Newtonian gauge. Breaking the gravitational equation of motion into pieces following Eq. \eqref{eq:masterEOM}, the $Y$ mode of the time-time component can be written as
\begin{align}
  \Mp \Omega(t) \left[
    \frac{2k^2 - 6k_0}{a^2} \phi^N
  + 6 H (\dot{\phi}^N + H \psi^N)
  \right]
  = - \delta \rho^N + \left.\tensor{T}{^0_0^{(Q)}}\right|_Y\,,
\end{align}
while the $Y_i$ mode of the time-space component is given by
\begin{align}
  \Mp \Omega(t) 2 k \left[ \dot{\phi}^N + H \psi^N \right] =
  (\bar{\rho}_m + \bar{P}_m) v^N
  + \left.\tensor{T}{^0_i^{(Q)}}\right|_{Y_i}.
\end{align}
Substituting the second of these equations into the first, we obtain the Poisson equation.
\begin{align}
  \Mp \Omega \frac{2k^2 - 6k_0}{a^2} \phi^N
  =
  - \bar{\rho}_m \Delta
  + \left.\tensor{T}{^0_0^{(Q)}}\right|_Y
  - \frac{3H}{k} \left.\tensor{T}{^0_i^{(Q)}}\right|_{Y_i} \label{eq:Poisson}
\end{align}
Here, we have defined
\begin{align}
  \bar{\rho}_m \Delta = \delta \rho^N
  + \frac{3H}{k} (\bar{\rho}_m + \bar{P}_m) v^N\,, \label{eq:Deltadef}
\end{align}
as the rest-frame density perturbation of matter.

The anisotropic shear stress is given by the $\tensor{Y}{^i_j}$ mode of the spatial components of the gravitational equation of motion. We express it in Newtonian gauge.
\begin{align}
  \Mp \Omega \frac{k^2}{a^2} \left(\phi^N - \psi^N\right)
  = \bar{P}_m \Pi
  + \left.\tensor{T}{^i_j^{(Q)}}\right|_{\tensor{Y}{^i_j}} \label{eq:shearstress}
\end{align}

In order to detail the contributions to the Poisson equation and anisotropic shear stress from our operators, we define $\Delta_P$ for an operator by its contribution to the Poisson equation by
\begin{align}
  \Mp \Omega \frac{2k^2 - 6k_0}{a^2} \phi^N
  =
  - \bar{\rho}_m \Delta
  + \ldots + \Delta_P.
\end{align}
We similarly define $\Delta_S$ for the anisotropic shear stress as
\begin{align}
  \Mp \Omega \frac{k^2}{a^2} \left(\phi^N - \psi^N\right)
  = \bar{P}_m \Pi + \ldots + \Delta_S.
\end{align}
In Table \ref{tab:poissonshear}, we present the contribution to the Poisson equation and anisotropic shear stress from the operators we considered. These quantities were extracted from the equations of motion, detailed in Appendix \ref{app:EOMs}.

\begin{table}
  \centering
  \small
  \noindent\makebox[\textwidth]{%
  \begin{tabular}{|>{$}c<{$}||>{$}c<{$}|>{$}c<{$}|}
    \hline
    \textbf{Operator} &
    \Delta_P & \Delta_S
    \\[1pt] \hline\hline
%
% Background terms
%
    \begin{aligned}
      \Omega(t), c(t), \Lambda(t)
    \end{aligned}
  &
    \begin{aligned}
      &- 2 c (\dot{\pi}^N - \psi^N) \\
      &+ \Mp \dot{\Omega} \left[
      - 3 \dot{H} \pi^N
      + \frac{k^2}{a^2} \pi^N
      - 3 \left(\dot{\phi}^N + H \psi^N \right)
  \right]
    \end{aligned}
  &
    \begin{aligned}
      \Mp \dot{\Omega} \frac{k^2}{a^2} \pi^N
    \end{aligned}
  \\[15pt] \hline
%
% (\delta g^{00})^2
%
    \begin{aligned}
      \frac{M_2^4 (t)}{2} (\delta g^{00})^2
    \end{aligned}
  &
    \begin{aligned}
      - 4 M_2^4 (\dot{\pi}^N - \psi^N)
    \end{aligned}
  &
    \begin{aligned}
      0
    \end{aligned}
  \\[6pt] \hline
%
% \delta g^{00} \delta \tensor{K}{^\mu_\mu}
%
    \begin{aligned}
      - \frac{\bar{M}_1^3 (t)}{2} \delta g^{00} \delta \tensor{K}{^\mu_\mu}
    \end{aligned}
  &
    \begin{aligned}
      - \bar{M}_1^3 \left(3 \dot{\phi}^N + 3 H \psi^N + 3 \dot{H} \pi^N - \frac{k^2}{a^2} \pi^N \right)
    \end{aligned}
  &
    \begin{aligned}
      0
    \end{aligned}
  \\[6pt] \hline
%
% (\delta \tensor{K}{^\mu_\mu})^2
%
    \begin{aligned}
      - \frac{\bar{M}_2^2 (t)}{2} (\delta \tensor{K}{^\mu_\mu})^2
    \end{aligned}
  &
    \begin{aligned}
      0
    \end{aligned}
  &
    \begin{aligned}
      0
    \end{aligned}
  \\[6pt] \hline
%
% \delta \tensor{K}{^\mu_\nu} \delta \tensor{K}{^\nu_\mu}
%
    \begin{aligned}
      - \frac{\bar{M}_3^2 (t)}{2} \delta \tensor{K}{^\mu_\nu} \delta \tensor{K}{^\nu_\mu}
    \end{aligned}
  &
    \begin{aligned}
      2 H \bar M_3^2 \frac{3 k_0 - k^2}{a^2} \pi^N
    \end{aligned}
  &
    \begin{aligned}
      -\bar M_3^2 \frac{k^2}{a^2} \left( \dot\pi^N + \Big(H+\frac{2\dot{\bar M}_3}{\bar M_3}\Big)\pi^N \right)
    \end{aligned}
  \\[8pt] \hline
%
% \delta g^{00} \delta R^{(3)}
%
    \begin{aligned}
      \frac{\hat{M}^2 (t)}{2} \delta g^{00} \delta R^{(3)}
    \end{aligned}
  &
    \begin{aligned}
       4 \hat{M}^2 \frac{3 k_0 - k^2}{a^2} (\phi^N + H \pi^N)
    \end{aligned}
  &
    \begin{aligned}
      2 \hat{M}^2 \frac{k^2}{a^2} (\psi^N - \dot{\pi}^N)
    \end{aligned}
  \\[6pt] \hline
%
% m_2^2 (t) (g^{\mu \nu} + n^\mu n^\nu) \partial_\mu g^{00} \partial_\nu g^{00}
%
    \begin{aligned}
      m_2^2 (t) \frac{\tilde{g}^{ij}}{a^2} \partial_i g^{00} \partial_j g^{00}
    \end{aligned}
  &
    \begin{aligned}
       8 m_2^2 \frac{k^2}{a^2} (\psi^N - \dot{\pi}^N)
    \end{aligned}
  &
    \begin{aligned}
      0
    \end{aligned}
  \\[6pt] \hline
  \end{tabular}}
  \normalsize
  \caption[Contributions to the Poisson equation and anisotropic shear stress]{
    The contributions to the Poisson equation and anisotropic shear stress from various operators. Note that for the Poisson equation contribution from the background operators, the background continuity equation \eqref{eq:darkcontinuity} has been used to simplify the result.
  }
\label{tab:poissonshear}
\end{table}

\subsection{Newtonian Limit}
In order to calculate the effective Newtonian constant, we need to take the Newtonian limit of our equations of motion. The Poisson equation, the anisotropic stress equation, and the equation of motion for $\pi$ contain the information we need. We follow the method described in Ref. \cite{DeFelice2011} closely.

In the Newtonian limit, we look at modes that are deep within the horizon\footnote{De Felice \textit{et al.} \cite{DeFelice2011} point out that if the speed of sound is much less than one, then the sound horizon should be used instead of the Hubble horizon, which may limit the regime of validity of this approximation substantially. For nonstandard dispersion relations, this notion must be generalized appropriately.}, $k^2/a^2 \gg H^2$. We also make the quasistatic approximation, discarding time derivatives of the fields. Finally, we assume that matter gives rise to no anisotropic shear stress.

To include the effect of possible perturbative terms, we write the Poisson equation, anisotropic stress equation, and $\pi$ equation of motion in this limit with arbitrary coefficients:
\begin{subequations}
\begin{align}
  A_1 \frac{k^2}{a^2} \phi^N + A_2 \frac{k^2}{a^2} \pi^N + A_3 \frac{k^2}{a^2} \psi^N &\simeq - \bar{\rho}_m \Delta \label{eq:newtlimit1}\,,
\\
  B_1 \psi^N + B_2 \phi^N + B_3 \pi^N &\simeq 0 \label{eq:newtlimit2}\,,
\\
  C_1 \frac{k^2}{a^2} \phi^N + C_2 \frac{k^2}{a^2} \psi^N + \left(C_3 \frac{k^2}{a^2} + M^2\right) \pi^N &\simeq 0\,. \label{eq:newtlimit3}
\end{align}
\end{subequations}
We keep the ``mass'' term $M^2$ in the $\pi$ equation of motion; whether or not this term is suppressed will depend on how it compares to the mode in question. Note that $M^2$ actually has mass dimension 6, because the $\pi$ field is not canonically normalized. In Table \ref{tab:coeffs}, we present the contribution to each coefficient from the operators we have considered. The overall sign and normalization for the coefficients are given as follows. The $A_i$ coefficients can be read off from $-\Delta_P$, while the $B_i$ coefficients can be read off from $- \Delta_S a^2 / k^2 \Mp \Omega$ (the normalization is chosen for simplicity). For the background contribution, the LHS of Eqs. \eqref{eq:Poisson} and \eqref{eq:shearstress} need to also be included. Contributions to the $C_i$ coefficients and $M^2$ can be read off from $\Delta_\pi$, the contribution of individual operators to the $\pi$ equation of motion, which are presented in Appendix \ref{app:EOMs}.

\begin{table}
  \centering
  \footnotesize
  \noindent\makebox[\textwidth]{%
  \begin{tabular}{|>{$}c<{$}||>{$}c<{$}|>{$}c<{$}|>{$}c<{$}|>{$}c<{$}|>{$}c<{$}|>{$}c<{$}|>{$}c<{$}|}
    \hline
    &
    \Omega, \Lambda, c &
    M_2^4 &
    \bar{M}_1^3 &
    \bar{M}_2^2 &
    \bar{M}_3^2 &
    \hat{M}^2 &
    m_2^2
    \\ \hline
    & &
    (\delta g^{00})^2 &
    \delta g^{00} \delta \tensor{K}{^\mu_\mu} &
    (\delta \tensor{K}{^\mu_\mu})^2 &
    \delta \tensor{K}{^\mu_\nu} \tensor{K}{^\nu_\mu} &
    \delta g^{00} \delta R^{(3)} &
    \frac{\tilde{g}^{ij}}{a^2} \partial_i g^{00} \partial_j g^{00}
    \\ \hline \hline
    A_1 & 2 \Mp \Omega & 0 & 0 & 0 & 0 & 4 \hat{M}^2 & 0
    \\ \hline
    A_2 & - \Mp \dot{\Omega} & 0 & - \bar{M}_1^3 & 0 & 2 H \bar M_3^2 & 4 H \hat{M}^2 & 0
    \\ \hline
    A_3 & 0 & 0 & 0 & 0 & 0 & 0 & - 8 m_2^2
    \\ \hline
    B_1 & -1 & 0 & 0 & 0 & 0 & - \frac{2 \hat{M}^2}{\Mp \Omega} & 0
    \\ \hline
    B_2 & 1 & 0 & 0 & 0 & 0 & 0 & 0
    \\ \hline
    B_3 & - \frac{\dot{\Omega}}{\Omega} & 0 & 0 & 0 & \frac{\bar M_3^2}{\Mp \Omega} \left(H+\frac{2\dot{\bar M}_3}{\bar M_3}\right) & 0 & 0
    \\ \hline
    C_1 & \Mp \dot{\Omega} & 0 & 0 & 0 & 0 & 2 H \hat{M}^2 + 4 \hat{M} \dot{\hat{M}} & 0
    \\ \hline
    C_2 & - \frac{\Mp}{2} \dot{\Omega} & 0 & -\frac{\bar{M}_1^3}{2} & - \frac{3}{2} \bar{M}_2^2 H & -\frac{\bar M_3^2}{2} H & 2 H \hat{M}^2 & 0
    \\ \hline
    C_3 & c & 0 & - \frac{1}{2} (H + \partial_t) \bar{M}_1^3 & \begin{aligned}&- 3 \bar{M}_2^2 \dot{H} \\&+ \frac{\bar{M}_2^2}{2} \frac{k^2}{a^2}\end{aligned} & \bar M_3^2 \left(\frac{k^2}{2a^2}-\dot H - \frac{k_0}{a^2}\right) & 2(H^2 + \dot{H} + H \partial_t) \hat{M}^2 & 0
    \\[12pt] \hline
    M^2 & \begin{aligned}&\frac{\Mp}{4} \dot{\Omega} \dot{R}^{(0)} \\ &- 3 c \dot{H} \end{aligned} & 0 & \begin{aligned}&\frac{3}{2} \dot{H} [(3H + \partial_t) \bar{M}_1^3] \\ &+ \frac{3}{2} \ddot{H} \bar{M}_1^3 \end{aligned} & \frac{9}{2} \bar{M}_2^2 \dot{H}^2 & \frac{3}{2}\bar M_3^2\dot H^2 & -6 \frac{k_0}{a^2} (H^2 + \dot{H} + H \partial_t) \hat{M}^2 & 0
    \\[12pt] \hline
  \end{tabular}}
  \normalsize
  \caption[Contributions to coefficients of equations in the Newtonian limit]{
    The contributions to the Poisson equation, anisotropic shear stress, and $\pi$ equations of motion in the Newtonian limit.
  }
\label{tab:coeffs}
\end{table}

To obtain the effective Newton constant, we want to form the equation
\begin{align}
  \frac{k^2}{a^2} \psi = - 4 \pi G_{\mathrm{eff}} \bar{\rho}_m \Delta.
\end{align}
Take Eq. \eqref{eq:newtlimit2}, solve it for $\phi^N$, and substitute this into Eqs. \eqref{eq:newtlimit1} and \eqref{eq:newtlimit3}. Then, solve \eqref{eq:newtlimit3} for $\pi^N$, substitute this into \eqref{eq:newtlimit1}, and solve for $G_{\mathrm{eff}}$.
\begin{align}
  4 \pi G_{\mathrm{eff}} =
  \frac{B_2 C_3 - C_1 B_3 + \frac{a^2}{k^2} B_2 M^2}
  {
    A_1 (B_3 C_2 - B_1 C_3 - B_1 M^2 a^2/k^2)
  + A_2 (B_1 C_1 - B_2 C_2)
  + A_3 (B_2 C_3 - B_3 C_1 + B_2 M^2 a^2/k^2)
  }
\end{align}

In general relativity, the result is
\begin{align}
  4 \pi G_{\mathrm{eff}} = \frac{1}{2 m_P^2}.
\end{align}
By including the background operators, this becomes
\begin{align}
  4 \pi G_{\mathrm{eff}} = \frac{1}{2 \Mp \Omega} \left( 1 +
      \frac{\Mp \dot{\Omega}^2}
      {
        4 \Omega c
      + 3 \Mp \dot{\Omega}^2
      + 4 \Omega \frac{a^2}{k^2} M^2
      }
      \right).
\end{align}
The expressions rapidly become more complicated as further terms are included. It is worth noting that of the terms we consider, only $\bar{M}_1$, $\bar{M}_3$, $\hat{M}$ and $m_2$ modify $G_{\mathrm{eff}}$ if $\dot{\Omega} = 0$.

We can also use Eqs. \eqref{eq:newtlimit2} and \eqref{eq:newtlimit3} to eliminate $\pi^N$ and solve for $\varpi = \psi^N/\phi^N - 1$, where $\varpi$ is defined by Caldwell \textit{et al.} \cite{caldwell2007} as the cosmological analog of Eddington's PPN $\gamma$ parameter ($\varpi \approx 1 - \gamma$).
\begin{align}
  \varpi =
    \frac
    {B_2 C_3 - C_1 B_3 + \frac{a^2}{k^2} B_2 M^2}
    {B_3 C_2 - B_1 C_3 - \frac{a^2}{k^2} B_1 M^2} - 1
\end{align}
This result for $\varpi$ is calculated from a quadratic action, and so we stress that nonlinear effects such as the Vainshtein \cite{Vainshtein1972, Luty2003,Alberto2004} and chameleon \cite{Khoury2004,Khoury2004a} mechanisms are not present. Thus, solar system constraints on $\gamma$ cannot be used to constrain $\varpi$. However, cosmological observations may be used to do so; Caldwell \textit{et al.} \cite{caldwell2007} place a bound of $\varpi = 0.02 \pm 0.07$ ($2 \sigma$) on tens of kiloparsec scales, and $\varpi = 0.03 \pm 0.10$ ($2 \sigma$) on hundreds of kiloparsec scales. Unfortunately, no useful limits were obtained on Megaparsec scales.

In general relativity, $\varpi = 0$. Looking at the background terms alone, $\varpi$ is given by
\begin{align}
  \varpi = \frac
      {\Mp \frac{\dot{\Omega}^2}{2 \Omega}}
      {c + \Mp \frac{\dot{\Omega}^2}{2 \Omega} + \frac{a^2}{k^2} M^2}.
\end{align}
Again, the expressions for $\varpi$ become rapidly more complicated as further terms are included. In the case where $\dot{\Omega} = 0$, only the $\bar{M}_3$ and $\hat{M}$ terms contribute, however.

\subsection{Phenomenological Functions}
In order to compare observational data with modification to general relativity, Bean and Tangmatitham \cite{bean2010} introduced two phenomenological functions $Q$ and $R$ which perturbed the Poisson equation and anisotropic shear stress. Here, we construct their functions from our formalism. Other phenomenological parameterizations have also been proposed; see \cite{bean2010} and references therein.

As we will be performing comparisons to general relativity in this section, we need to decide what form Newton's constant $G$ will take in the comparisons. For simplicity, we choose to scale $\Omega$ such that $\Mp = 1/8 \pi G$, with $\Omega = 1$ today.

In general relativity, the Poisson equation is
\begin{align}
  (k^2 - 3k_0) \phi^N
  = {}& - 4 \pi G a^2 \bar{\rho}_m \Delta\,,
\end{align}
where $\Delta$ is defined in Eq. \eqref{eq:Deltadef}, while the anisotropic shear stress is given by\footnote{An alternative representation of the matter anisotropic shear (such as used by Ma and Bertschinger \cite{ma1995}) is given by $\sigma = 2 \bar{P}_m \Pi/3(\bar{\rho}_m + \bar{P}_m)$.}
\begin{align}
  \phi^N - \psi^N
  = 8 \pi G a^2 \frac{\bar{P}_m \Pi}{k^2}.
\end{align}
Bean and Tangmatitham introduce two phenomenological functions $Q$ and $R$ to describe deviations from general relativity by
\begin{align}
  (k^2 - 3k_0) \phi^N
  = {}& - 4 \pi G Q a^2 \bar{\rho}_m \Delta\,,
\end{align}
and
\begin{align}
  \phi^N - R \psi^N
  = 8 \pi G a^2 \frac{\bar{P}_m \Pi}{k^2}.
\end{align}

We may write our Poisson equation \eqref{eq:Poisson} in this form as
\begin{align}
  (k^2 - 3k_0) \phi^N
  =
  - 4 \pi G a^2 \bar{\rho}_m \Delta \left[ 1 + \frac{1}{\bar{\rho}_m \Delta} \left(
  - \Delta_P
  + m_P^2 (\Omega - 1) \frac{2 k^2 - 6 k_0}{a^2} \phi^N \right)
  \right] \,,
\end{align}
and thus identify
\begin{align}
  Q = 1 + \frac{1}{\bar{\rho}_m \Delta} \left(
  - \Delta_P
  + m_P^2 (\Omega - 1) \frac{2 k^2 - 6 k_0}{a^2} \phi^N \right) \,,
\end{align}
where the sum is intended to indicate the combined contribution over all operators. Similarly, we can write our equation for the anisotropic shear stress \eqref{eq:shearstress} as
\begin{align}
  \phi^N - \psi^N \left( 1
  - (\Omega - 1) \left(\frac{\phi^N}{\psi^N} - 1\right)
  + \frac{1}{\psi^N} \frac{a^2}{m_P^2 k^2} \Delta_S \right)
  = 8 \pi G a^2 \frac{\bar{P}_m \Pi}{k^2}\,,
\end{align}
leading to
\begin{align}
  R = 1
  - (\Omega - 1) \left(\frac{\phi^N}{\psi^N} - 1\right)
  + \frac{1}{\psi^N} \frac{a^2}{m_P^2 k^2} \Delta_S.
\end{align}

For a typical dark energy model, the function $Q$ is expected to remain close to unity until the very late stages of the universe, as the energy density in dark energy perturbations will be dwarfed by that of the matter perturbations. The $R$ parameter is more interesting, and displays the possibility of deviating from unity much earlier, as it is the ratio of the dark energy perturbations to the gravitational perturbations which is important. The non-minimal coupling $\Omega$ enhances (for $\Omega > 1$) the effect of existing gravitational potentials and anisotropic shear stresses.

Bean and Tangmatitham found that a constant $Q$ and $R$ were consistent with unity at the 95\% confidence level. However, for $Q$ and $R$ that became important in the late universe, much more freedom was available, particularly on large scales. At the time that work was performed, observational data was insufficient to lift a degeneracy between $Q$ and $R$. It is hoped that newer data will be able to lift this degeneracy.

\subsection{Confronting Models with Observations}

Having calculated the contribution from various operators in our EFT action to different observables, we can now combine this with our knowledge of different dark energy models. Table \ref{tab:models} summarizes the operators required to describe various models.

\begin{table}[ht]
  \centering
  \small
  \noindent\makebox[\textwidth]{%
  \begin{tabular}{|p{3.45cm}||c|c|c|c|c|c|c|c|c|}
    \hline
    \textbf{Operator} &
    $\Omega$ &
    $\Lambda$ &
    $c$ &
    $M_2^4$ &
    $\bar{M}_1^3$ &
    $\bar{M}_2^2$ &
    $\bar{M}_3^2$ &
    $\hat{M}^2$ &
    $m_2^2$
    \\ \hline
    \textbf{Model} &
    R &
    &
    $\delta g^{00}$ &
    $(\delta g^{00})^2$ &
    $\delta g^{00} \delta \tensor{K}{^\mu_\mu}$ &
    $(\delta \tensor{K}{^\mu_\mu})^2$ &
    $\delta \tensor{K}{^\mu_\nu} \tensor{K}{^\nu_\mu}$ &
    $\delta g^{00} \delta R^{(3)}$ &
    $\frac{\tilde{g}^{ij}}{a^2} \partial_i g^{00} \partial_j g^{00}$
    \\ \hline
    \hline
    $\Lambda$CDM & 1 & \checkmark & 0 & - & - & - & - & - & - \\ \hline
    Quintessence & 1/\checkmark & \checkmark & \checkmark & - & - & - & - & - & - \\ \hline
    $F(R)$ & \checkmark & \checkmark & 0 & - & - & - & - & - & - \\ \hline
    $k$-essence & 1/\checkmark & \checkmark & \checkmark & \checkmark & - & - & - & - & - \\ \hline
    Galileon \cite{Chow:2009fm} \newline Kinetic Braiding \cite{Deffayet:2010qz} & 1/\checkmark & \checkmark & \checkmark & \checkmark & \checkmark & - & - & - & - \\ \hline
    DGP \cite{Dvali2000} & \checkmark & \checkmark $\dagger$ & \checkmark $\dagger$ & \checkmark $\dagger$ & \checkmark & - & - & - & - \\ \hline
    Ghost Condensate \cite{Arkani-Hamed2004} & 1/\checkmark & \checkmark & 0 & - & - & \checkmark & \checkmark & - & - \\ \hline
    Horndeski \cite{Horndeski1974} & \checkmark & \checkmark & \checkmark & \checkmark & \checkmark & \checkmark $\dagger$ & \checkmark $\dagger$ & \checkmark $\dagger$ & - \\ \hline
    Ho\v{r}ava-Lifshitz \cite{Horava:2009uw} & 1 & \checkmark & 0 & - & - & \checkmark & - & - & \checkmark \\ \hline
  \end{tabular}}
  \normalsize
  \caption[List of operators associated with various models]{
    A list of operators required to describe various different models. The correspondences are explicitly presented in Refs. \cite{Gubitosi2012, Bloomfield2013}.\\
    \begin{tabular}{p{2cm}cp{0.7 \textwidth}}
      &\checkmark & Operator is necessary\\
      &- & Operator is not included\\
      &1, 0 & Coefficient is unity or exactly vanishing\\
      &1/\checkmark & Minimally and non-minimally coupled versions of this model exist\\
      &$\dagger$ & Coefficients marked with a dagger are linearly related to other coefficients in that model by numerical coefficients
    \end{tabular}
  }
\label{tab:models}
\end{table}

We have calculated three main quantities. The first is the speed of sound, which is closely related to stability considerations, and also to the clumping of dark energy. The dispersion relationship may also be modified away from the relativistic $\omega = c_s k$. The second is the modified Poisson equation in the Newtonian limit, which is important in describing the growth of structure. We describe the modified Poisson equation in terms of the effective Newtonian constant. The third is the effective anisotropic shear stress, described by Caldwell's $\varpi$ parameter, which appears in weak lensing equations. Each operator gives a different contribution to each of these quantities.

Before summarizing the effect of each operator on these quantities, we should begin by understanding what is predicted by the $\Lambda$CDM and basic quintessence models, so we know what we are comparing to. In $\Lambda$CDM, dark energy exists in the form of a cosmological constant. There is no speed of sound and no clumping of dark energy. The effective Newtonian constant is exactly the Newtonian constant $G_N$, and anisotropic shear stresses arise only from photons and neutrinos, which are of negligible importance in the late universe.

In contrast, a minimally coupled quintessence model has nonzero $\Lambda(t)$ and $c(t)$, and a speed of sound equal to the speed of light. The effective Newtonian constant is once again $G_N$ (in the Newtonian limit). There are no additional sources of anisotropic shear stress. As additional operators are turned on, we begin to see new effects appearing. We summarize these effects in Table \ref{tab:effects}.

By combining the data in Tables \ref{tab:models} and \ref{tab:effects}, we can predict which models will manifest different effects, and use this information to discriminate between different classes of models. It is hoped that with the help of numerical simulations, the coefficients of different operators can be constrained, leading to an overall picture of which classes of models yield a viable description of the growth of large scale structure.

\begin{table}[ht]
  \centering
  \small
  \noindent\makebox[\textwidth]{%
  \begin{tabular}{|p{4.75cm}||c|c|c|c|c|c|c|c|c|}
    \hline
    \textbf{Operator} &
    $\Omega(t)$ &
    $M_2^4$ &
    $\bar{M}_1^3$ &
    $\bar{M}_2^2$ &
    $\bar{M}_3^2$ &
    $\hat{M}^2$ &
    $m_2^2$
    \\ \hline
    \textbf{Observable} &
    R &
    $(\delta g^{00})^2$ &
    $\delta g^{00} \delta \tensor{K}{^\mu_\mu}$ &
    $(\delta \tensor{K}{^\mu_\mu})^2$ &
    $\delta \tensor{K}{^\mu_\nu} \tensor{K}{^\nu_\mu}$ &
    $\delta g^{00} \delta R^{(3)}$ &
    $\frac{\tilde{g}^{ij}}{a^2} \partial_i g^{00} \partial_j g^{00}$
    \\ \hline
    \hline
    Speed of Sound & 1 & \checkmark & \checkmark & \checkmark + $k^4$ & \checkmark + $k^4$ & \checkmark & $\ast$ \\ \hline
    Effective Newtonian Constant & \checkmark & - & \checkmark & \checkmark + $k^2$ & \checkmark + $k^2$ & \checkmark & \checkmark \\ \hline
    Caldwell's $\varpi$ parameter & \checkmark & - & \checkmark & \checkmark & \checkmark & \checkmark & - \\ \hline
  \end{tabular}}
  \normalsize
  \caption[Effects of operators on physical quantities]{
    The effect of various operators on physical quantities.\\
    \begin{tabular}{p{2cm}cp{0.7 \textwidth}}
      &\checkmark & Operator modifies this quantity\\
      &- & Operator does not modify this quantity\\
      &1 & Speed of sound is unity\\
      & \checkmark + $k^n$ & Operator introduces a new scale dependence\\
      & $\ast$ & Operator behaves unusually (see above discussion on speed of sound)
    \end{tabular}
  }
\label{tab:effects}
\end{table}

\section{Conclusions and Outlook}
We have proposed an effective field theory construction to describe dark energy and modified gravity models in the late universe. The construction of the action \eqref{eq:theaction} is performed in a unitary gauge, and consists of operators written in terms of deviations from an FRW background of quantities that are invariant under spatial coordinate transformations. Three of these EFT terms, along with the matter action, contribute to the background cosmological evolution. A number of further operators deemed interesting were identified.

Explicit scalar perturbations in our model were introduced in Section \ref{sec:pi}, where we also raised theoretical issues associated with the use of this construction at high redshifts. A systematic investigation into the physical effects of various operators was undertaken in Section \ref{sec:observations}, where we highlighted the importance of perturbative effects in constraining dark energy and modified gravity models.

On the theoretical side, further work towards understanding the details of the formalism at high redshifts is necessary. Following this, a detailed investigation into which operators are dominant at different energy scales and cosmological epochs would also be of interest. These calculations are complicated by the presence of both the matter fields and the non-minimal coupling $\Omega$, and thus results from inflationary work using this formalism are non-trivial to translate.

In addition, it will be important to determine the region
in parameter space in which ghosts, instabilities, and strong coupling
do not occur, for both the scalar and tensor sectors, in order
to insure a theoretically consistent framework.  It is straightforward
to analyze these issues for all the operators we have considered, one
at a time, but it is more complex to analyze all the operators
simultaneously.  We leave analysis of these important issues for
future work.

On the numerical side, this formalism represents an exciting opportunity to place model-independent constraints on dark energy and modified gravity models. Each individual operator leads to a characteristic physical signature, and follow-up work to constrain the behavior of the coefficients of the operators investigated here using observational data is underway \cite{Mueller2013}.

This formalism will also be of use in investigating explicit models of dark energy, by linking the behavior of perturbations in specific models to the parameters which also control their background evolution. As demonstrated in Table \ref{tab:models}, the matching to this formalism for a variety of dark energy and modified gravity models has already been performed. Observational constraints on individual models, for which the form of the coefficients is predicted, should be stronger than the general bounds that can be set from model-independent analyses.

%%%%%%%%%%%%%%%%%%%%%%%%%%%%%%%%%%%%%%%%%%%%%%%%%%%
%%%%%%%%%%%%%%%%%%%%%%%%%%%%%%%%%%%%%%%%%%%%%%%%%%%
%%%%%%%%%%%%%%%%%%%%%%%%%%%%%%%%%%%%%%%%%%%%%%%%%%%

\section*{Acknowledgments}
\vspace{-0.2in}
We thank Rachel Bean, Sera Cremonini, Liam McAllister
and Eva-Maria Mueller for useful conversations. SW would like to thank
the George
and Cynthia Mitchell Institute for Fundamental Physics, Texas A\&M
University for hospitality. MP is partially supported by the
U.S. National Science Foundation grant PHY-1205986. EF and JB were
supported in part by NSF grants PHY-1068541 and PHY-0968820, by NASA
grants NNX11AI95G, and by the John and David Boochever Prize
Fellowship in Fundamental Theoretical Physics to JB at Cornell.

\appendix

%%%%%%%%%%%%%%%%%%%%%%%%%%%%%%%%%%%%%%%%%%%%%%%%%%
%%%%%%%%%%%%%%%%%%%%%%%%%%%%%%%%%%%%%%%%%%%%%%%%%%
%%%%%%%%%%%%%%%%%%%%%%%%%%%%%%%%%%%%%%%%%%%%%%%%%%

\section{Applying the St\"uckelberg Trick} \label{app:goldstone}

In this appendix, we give the transformation law for various quantities using the St\"uckelberg trick to introduce the $\pi$ field.

Functions of time transform as $t \rightarrow t + \pi$.
\begin{align}
  f(t) \rightarrow f(t + \pi) \approx f(t) + \dot{f}(t) \pi + \frac{\ddot{f}(t)}{2} \pi^2 + \ldots
\end{align}
Such functions will typically be Taylor-expanded about the background. For terms that start at second order in perturbations, we ignore the expansion of their time-dependent coefficients, under the assumption that they are not rapidly oscillating.

Operators that are diffeomorphism invariant do not receive any contributions from $\pi$. For example, the Ricci scalar $R$. However, when looking at the perturbative terms, e.g. $\delta R$, the functions of time forming the background part of the term need to be expanded. For example,
\begin{align}
  \delta R = R - R^{(0)} \rightarrow R - R^{(0)} - \dot{R}^{(0)} \pi - \ldots \,.
\end{align}
Care must be taken when dealing with the normal and extrinsic curvature tensor however, as these quantities depend on the foliation, which also changes under a transformation.

Any tensor that includes explicit time components in the action must transform using the tensor transformation law. For example,
\begin{align}
  f^0 &\rightarrow \frac{\partial (t + \pi(x^\lambda))}{\partial x^\mu} f^{\mu}
  = f^0 + \partial_{\mu} \pi f^{\mu}.
\end{align}
We will often need to transform the time-time component of the inverse metric, $g^{00}$ (or its perturbation, which transforms in the same way). It transforms as
\begin{align}
  g^{00} &\rightarrow \frac{\partial (t + \pi(x^\lambda))}{\partial x^\mu} \frac{\partial (t + \pi(x^\lambda))}{\partial x^\nu} g^{\mu \nu}
  = g^{00} - 2 \dot{\pi} + 2 \dot{\pi} \delta g^{00} + 2 \tilde{\nabla}_i \pi g^{0i} - \dot{\pi}^2 + \frac{(\tilde{\nabla} \pi)^2}{a^2},
\end{align}
where we have truncated at second order in perturbations. We will often only need this expansion to first order.

The normal doesn't just follow the tensor transformation law, as it also depends on the foliation of spacetime, which changes with the coordinate transformation. Our foliation is initially described by $t = \mathrm{const}$. The normal to these surfaces is given by
\begin{align}
  n_\mu = \frac{\partial_\mu t}{\sqrt{-g^{\mu \nu} \partial_\mu t \partial_\nu t}}.
\end{align}
On the background, this reduces to $n_\mu^{(0)} = \delta^0_\mu$.

When the $\pi$ field is introduced, we have $t = \tilde{t} + \pi$, $x = \tilde{x}$. The index on the normal transforms following the tensor transformation law, while the normal itself transforms because of the changing foliation.
\begin{align}
  n_{\nu} = \frac{\partial \tilde{x}^{\mu}}{\partial x^\nu} \frac{\partial_{\mu} (\tilde{t} + \pi)}{\sqrt{-g^{\alpha \beta} \partial_{\alpha} (\tilde{t} + \pi) \partial_{\beta} (\tilde{t} + \pi)}}
\end{align}
Expanded to linear order in $\pi$ (and ignoring metric perturbations), this gives
\begin{align}
  n_{\nu} &= \frac{\partial \tilde{x}^{\mu}}{\partial x^\nu} \left[
  \tilde{n}_{\mu}
  \left(
  1 - \dot{\pi}
  \right)
  +
  \partial_{\mu} \pi
  \right]\,,
\end{align}
where $\tilde{n}_{\mu}$ is the normal to slices of constant time $\tilde{t}$.

For the extrinsic curvature tensor, we use the following definition.
\begin{align}
  \tensor{K}{^\mu_\nu} = (g^{\mu \sigma} + n^\mu n^\sigma) \nabla_\sigma n_\nu
\end{align}
Because $\tensor{K}{^\mu_\nu} n^\nu = \tensor{K}{^\mu_\nu} n_\mu = 0$, a zero component for the extrinsic curvature tensor will never explicitly appear in the unitary gauge action. Thus, the indices for this object will always be contracted, and we only need to concern ourselves with the transformation that arises from the change to the foliation (and not from the tensor transformation law, which will cancel from the transformation of the corresponding contracted index). We use the first order transformation of the normal from above. Note that to linear order, $n_0 \rightarrow \tilde{n}_0$ and $n_i \rightarrow \tilde{n}_i + \partial_i \pi$, so we should look at time and space components separately.

The extrinsic curvature tensor transforms as follows, to linear order, where we ignore the tensor transformation law and drop tildes.
\begin{align}
  \tensor{K}{^\sigma_\mu} \rightarrow
  \tensor{K}{^\sigma_\mu}
  +
  \left(g^{\sigma \lambda} + n^{\sigma} n^{\lambda} \right) \nabla_\lambda \left(- n_{\mu} \dot{\pi} + \nabla_{\mu} \pi \right)
  +
  \left(\nabla^{\sigma} \pi - n^{\sigma} \dot{\pi} \right) n^{\lambda} \nabla_\lambda n_{\mu}
  +
  n^{\sigma} \left(\nabla^{\lambda} \pi - n^{\lambda} \dot{\pi} \right) \nabla_\lambda n_{\mu}
\end{align}
In terms of individual space and time components, this transforms as
\begin{align}
  \tensor{K}{^0_0} &\rightarrow
  \tensor{K}{^0_0}\,,
\\
  \tensor{K}{^0_i} &\rightarrow
  \tensor{K}{^0_i} + H \tilde{\nabla}_i \pi\,,
\\
  \tensor{K}{^i_0} &\rightarrow
  \tensor{K}{^i_0}
  - H \frac{\tilde{g}^{ij}}{a^2} \tilde{\nabla}_j \pi\,,
\\
  \tensor{K}{^i_j} &\rightarrow
  \tensor{K}{^i_j}
  + \frac{\tilde{g}^{i k}}{a^2} \tilde{\nabla}_k \tilde{\nabla}_j \pi\,.
\end{align}
The trace of the extrinsic curvature tensor transforms as
\begin{align}
  \tensor{K}{^\mu_\mu} \rightarrow \tensor{K}{^\mu_\mu} + \frac{\tilde{\nabla}^2 \pi}{a^2}.
\end{align}

The Ricci scalar of the spatial metric is diffeomorphism invariant under (time-dependent) spatial transformations, but not under time diffeomorphisms. That this is the case may be seen from the contracted Gauss-Codazzi relation
\begin{align}
  R^{(3)} = R + 2 R^{\mu \nu} n_\mu n_\nu - K^2 + K_{\mu \nu} K^{\mu \nu}
\end{align}
where the RHS transforms nontrivially. We can use this relation to calculate the transformation of the three-dimensional Ricci scalar.
\begin{align}
  R^{(3)} \rightarrow R^{(3)} + 4 H \frac{\tilde{\nabla}^2 \pi}{a^2}
\end{align}

%%%%%%%%%%%%%%%%%%%%%%%%%%%%%%%%%%%%%%%%%%%%%%%%%%
%%%%%%%%%%%%%%%%%%%%%%%%%%%%%%%%%%%%%%%%%%%%%%%%%%
%%%%%%%%%%%%%%%%%%%%%%%%%%%%%%%%%%%%%%%%%%%%%%%%%%

\section{Metrics and Gauges}\label{app:Synchronous}

To remain general with respect to the choice of spatial coordinates, we do not specify a spatial metric. This is important for working with spatial metrics involving spatial curvature. We write the background FRW metric as
\begin{align}
  ds^2 = -dt^2 + a(t)^2 \tilde{g}_{ij} dx^i dx^j\,,
\end{align}
where $\tilde{g}_{ij}$ is a three-dimensional spatial metric with no time dependence. To ensure that this describes an FRW metric, we require that
\begin{align}
  \tilde{R}_{ijkl} = k_0 \left(\tilde{g}_{ik} \tilde{g}_{jl} - \tilde{g}_{il} \tilde{g}_{jk} \right)\,,
\end{align}
where $k_0$ is the curvature constant (we reserve $k$ for wavenumber, and use $K$ for the trace of the extrinsic curvature tensor). The metric determinant is
\begin{align}
  \sqrt{-g} = a^3 \sqrt{\tilde{g}}.
\end{align}

For metric perturbations, we perform most of our calculations in synchronous gauge, and transform the results to Newtonian gauge for generality. We consider perturbations of the form
\begin{align}
  ds^2 = -dt^2 + a(t)^2 (\tilde{g}_{ij} + h_{ij}) dx^i dx^j. \label{eq:pertmetric}
\end{align}
There are two scalar degrees of freedom in $h_{ij}$. The first is defined as the trace $h = \tilde{g}^{ij} h_{ij} = a^2 g^{ij} h_{ij}$. The second is $\eta$, defined by
\begin{align}
  h_{ij} = \frac{h}{3} \tilde{g}_{ij} + \left( \tilde{\nabla}_i \tilde{\nabla}_j - \frac{\tilde{g}_{ij}}{3} \tilde{g}^{kl} \tilde{\nabla}_k \tilde{\nabla}_l \right) \eta\,,
\end{align}
where $\tilde{\nabla}$ is the covariant derivative associated with $\tilde{g}_{ij}$. As a purely spatial tensor, we will raise and lower indices in $h_{ij}$ with the spatial metric $\tilde{g}_{ij}$ and its inverse only.

The connection coefficients can be written to first order in $h_{ij}$ as
\begin{align}
  \tensor{\Gamma}{^t_{tt}} &= \tensor{\Gamma}{^t_{ti}} = \tensor{\Gamma}{^i_{tt}} = 0\,,
\\
  \tensor{\Gamma}{^t_{ij}} &= H g_{ij} + H a^2 h_{ij} + \frac{1}{2} a^2 \dot{h}_{ij}\,,
\\
  \tensor{\Gamma}{^i_{jt}} &= H \delta^i_j + \frac{1}{2} \tensor{\dot{h}}{^i_j}\,,
\\
  \tensor{\Gamma}{^i_{jk}} &= \tensor{\tilde{\Gamma}}{^i_{jk}} + \frac{1}{2} \tilde{g}^{il} \left[ \tilde{\nabla}_k h_{jl} + \tilde{\nabla}_j h_{kl} - \tilde{\nabla}_l h_{jk} \right]\,.
\end{align}
The coefficient $\tensor{\Gamma}{^i_{jk}}$ is written in terms of the spatial metric and its connection $\tensor{\tilde{\Gamma}}{^i_{jk}}$, which is also used by the spatial covariant derivative $\tilde{\nabla}_k$. Note that spatial covariant derivatives commute with time derivatives.

Using these connection coefficients, the d'Alembertian of a function of time becomes
\begin{align}
  \square f =
  - \ddot{f}
  - \left(3 H + \frac{1}{2} \dot{h}\right) \dot{f}\,,
\end{align}
while the d'Alembertian of an arbitrary function to leading order is
\begin{align}
  \square f =
  - \ddot{f}
  - 3 H \dot{f}
  + \frac{1}{a^2} \tilde{\nabla}^2 f\,,
\end{align}
where $\tilde{\nabla}^2$ represents the spatial Laplacian.

The curvature invariants to first order in scalar perturbations are given below. We do not provide the Riemann tensor here, instead referring readers to Appendix D of Kodama and Sasaki \cite{kodama1984}. The Ricci tensor is as follows.
\begin{align}
  R_{t t} = {}& - 3 \left(\dot{H} + H^2 \right) - H \dot{h} - \frac{1}{2} \ddot{h}
\\
  R_{it} = {}& \frac{1}{2} \tilde{\nabla}_j \tensor{\dot{h}}{^j_i} - \frac{1}{2} \tilde{\nabla}_i \dot{h}
  = \tilde{\nabla}_i  \left( k_0 \dot{\eta} - \frac{1}{3} \dot{h} + \frac{1}{3} \tilde{\nabla}^2 \dot{\eta} \right)
\\
  R_{ij} = {}&
  \frac{2 k_0}{a^2} g_{ij} + \left(\dot{H} + 3 H^2\right) \left(g_{ij} + a^2 h_{ij}\right) + \frac{3}{2} H a^2 \dot{h}_{ij} + \frac{1}{2} a^2 \ddot{h}_{ij}
  + \frac{1}{2} \dot{h} H g_{ij}
\nonumber\\&
  + \frac{1}{6} \left(\tilde{\nabla}_i \tilde{\nabla}_j + \tilde{g}_{ij} \tilde{\nabla}^2 \right) \left(
  \tilde{\nabla}^2 \eta
  - h
  \right)
  + 2 k_0 \tilde{\nabla}_i \tilde{\nabla}_j \eta
\end{align}
Combining these components to calculate the Ricci scalar yields
\begin{align}
  R &= 6 \left(\dot{H} + 2 H^2 + \frac{k_0}{a^2}\right) + 4 H \dot{h} + \ddot{h} - 2 \frac{k_0}{a^2} h + \frac{1}{a^2}\left(
  2 k_0 \tilde{\nabla}^2 \eta
  - \frac{2}{3} \tilde{\nabla}^2 h
  + \frac{2}{3} \tilde{\nabla}^2 \tilde{\nabla}^2 \eta
  \right).
\end{align}
The Einstein tensor becomes as follows.
\begin{align}
  G_{tt} ={}& 3 \left( H^2 + \frac{k_0}{a^2}\right)
  + H \dot{h}
  - \frac{k_0}{a^2} h
  + \frac{1}{a^2}\left(
  k_0 \tilde{\nabla}^2 \eta
  - \frac{1}{3} \tilde{\nabla}^2 h
  + \frac{1}{3} \tilde{\nabla}^2 \tilde{\nabla}^2 \eta
  \right)
\\
  G_{it} ={}& R_{it}
\\
  G_{ij} ={}& - g_{ij} \left[ 3 H^2 + 2 \dot{H} + \frac{k_0}{a^2} \right] - a^2 h_{ij} \left(3 H^2 + 2 \dot{H} + 3 \frac{k_0}{a^2}\right)
\nonumber\\&
  + \frac{3}{2} H a^2 \dot{h}_{ij} + \frac{1}{2} a^2 \ddot{h}_{ij}
  - \frac{3}{2} \dot{h} H g_{ij} - g_{ij} \frac{1}{2} \ddot{h} + g_{ij} \frac{k_0}{a^2} h
\nonumber\\&
  + \frac{1}{6} \tilde{\nabla}_i \tilde{\nabla}_j \left( \tilde{\nabla}^2 \eta - h \right)
  - \frac{1}{6 a^2} g_{ij} \tilde{\nabla}^2 \left( \tilde{\nabla}^2 \eta - h \right)
  + 2 k_0 \tilde{\nabla}_i \tilde{\nabla}_j \eta
  - k_0 \frac{g_{ij} }{a^2} \tilde{\nabla}^2 \eta
\end{align}

The normal to surfaces of constant time is particularly straightforward in synchronous gauge.
\begin{align}
  n_\mu = \frac{\partial_\mu t}{\sqrt{- g^{\alpha \beta} \partial_\alpha t \partial_\beta t}} = \delta^0_\mu
\end{align}
We calculate the extrinsic curvature tensor from the normal. We use the definition
\begin{align}
  K_{\mu \nu} = \left(\delta_\mu^\sigma + n_\mu n^\sigma \right) \nabla_\sigma n_\nu.
\end{align}
To first order in perturbations, the components are
\begin{align}
  K_{00} = K_{i0} = 0
  , \qquad \qquad
  K_{ij} = - \tensor{\Gamma}{^t_{ij}} = - H g_{ij} - H a^2 h_{ij} - \frac{1}{2} a^2 \dot{h}_{ij}.
\end{align}
The trace of the extrinsic curvature $K = g^{\mu \nu} K_{\mu \nu}$ is then given by
\begin{align}
  K = - 3 H - \frac{1}{2} \dot{h}.
\end{align}

\subsection{Gauge Transformations}
For some calculations, particularly the anisotropic shear stress and the effective Poisson equation, it is convenient to work in Newtonian gauge. The transformations from synchronous to Newtonian gauge are well known. See, for example, Kodama and Sasaki \cite{kodama1984} or Ma and Bertschinger \cite{ma1995}. Here we provide the formulas for the transformation between synchronous gauge and Newtonian gauge. We focus only on the scalar modes of interest.

Begin with a general metric with perturbations written as
\begin{align}
  ds^2 = -(1 + 2 \psi) dt^2 + a(t)^2 \left[(1 - 2 \phi)\tilde{g}_{ij} + \chi_{ij}\right] dx^i dx^j\,,
\end{align}
where $\chi_{ij}$ is traceless. The special case of $\psi = 0$ defines synchronous gauge, and $\chi_{ij} = 0$ defines Newtonian gauge. Perform a coordinate transformation $x^\mu \rightarrow \hat{x}^\mu = x^\mu + \epsilon^\mu$, where we raise and lower indices on $\epsilon^\mu$ using the background metric. We choose $\epsilon^\mu$ to be
\begin{align}
  \epsilon_i = a^2 \nabla_i \beta
, \qquad
  \epsilon_0 = - a^2 \dot{\beta}\,,
\end{align}
for some scalar function $\beta$. This transformation preserves the form of the perturbed metric. The perturbations transform as follows, where we use hats to denote the new quantities.
\begin{align}
  \hat{\psi} &= \psi - a^2 \ddot{\beta} - 2 a^2 H \dot{\beta}
\\
  \hat{\phi} &= \phi + a^2 H \dot{\beta} + \frac{1}{3} \tilde{\nabla}^2 \beta
\\
  \hat{\chi}_{ij} &= \chi_{ij} - 2 \left( \tilde{\nabla}_i \tilde{\nabla}_j - \frac{1}{3} \tilde{g}_{ij} \tilde{\nabla}^2 \right) \beta
\end{align}

To transform from synchronous to Newtonian gauge, choose $\beta = \eta / 2$. The perturbations in Newtonian gauge are then as follows.
\begin{align}
  \psi^N &= - \frac{1}{2} a^2 \ddot{\eta} - a^2 H \dot{\eta}
\\
  \phi^N &= - \frac{h}{6} + \frac{1}{2} a^2 H \dot{\eta} + \frac{1}{6} \tilde{\nabla}^2 \eta
\\
  \chi^N_{ij} &= 0
\end{align}
When changing gauge, we must also remember that the $\pi$ field transforms, as described in Appendix \ref{app:goldstone}. Under the transformation described above, we have $\hat{\pi} = \pi - a^2 \dot{\beta}$, which becomes
\begin{align}
  \pi^N = \pi^S - \frac{a^2}{2} \dot{\eta}\,,
\end{align}
when transforming from synchronous gauge to Newtonian gauge.

When transforming tensors, the tensor transformation law also applies. For example, the Einstein tensor transforms as the following to linear order.
\begin{align}
  \tensor{\hat{G}}{^\mu_\nu} = \tensor{G}{^\mu_\nu} + \tensor{G}{^\alpha_\nu} \tensor{\epsilon}{^\mu_{,\alpha}} - \tensor{G}{^\mu_\alpha} \tensor{\epsilon}{^\alpha_{, \nu}} - \tensor{G}{^\mu_\nu_{,\alpha}} \epsilon^\alpha
\end{align}

In momentum space, the perturbations in synchronous and Newtonian gauge are related by
\begin{align}
  \psi^N &= \frac{a^2}{k^2} \left[\frac{1}{2} (\ddot{h} + 6 \ddot{\tilde{\eta}}) + H (\dot{h} + 6 \dot{\tilde{\eta}}) \right]\,,
\\
  \phi^N &= \tilde{\eta} - \frac{a^2 H}{2 k^2} \left[\dot{h} + 6 \dot{\tilde{\eta}}\right]\,,
\\
  \pi^N &= \pi^S + \frac{a^2}{2k^2} \left(\dot{h} + 6 \dot{\tilde{\eta}}\right) \,.
\end{align}
Note that these equations use $\tilde{\eta}$ rather than $\eta$. It is also useful to note that $\dot{\pi}^S = \dot{\pi}^N - \psi^N$, and $\tilde{\eta} + H \pi^S = \phi^N + H \pi^N$.

\section{Momentum Space} \label{app:momentum}
In this appendix, we detail the formalism we will use for the transformation to momentum space. This formalism is necessary for working with a metric involving spatial curvature.

We follow the conventions of Kodama and Sasaki \cite{kodama1984}. For each $\vec{k}$, define $Y_{\vec{k}}(x^i)$ to be a solution of the equation
\begin{align}
  \tilde{\nabla}^2 Y_{\vec{k}} = - k^2 Y_{\vec{k}}.
\end{align}
From now onwards, we suppress the $\vec{k}$ dependence of $Y$. Both $Y$ and $Y^*$ will be solutions of this equation, as will any linear combination. We fix this freedom by choosing
\begin{align}
  \lim_{k_0 \rightarrow 0} Y(x^i) =  e^{i \vec{k} \cdot \vec{x}}\,,
\end{align}
so that the modes become the usual Fourier modes when the background spatial metric is flat. We choose the normalization of $Y$ such that
\begin{align}
  \int d^3 x \sqrt{\tilde{g}} Y_{\vec{k}} Y^*_{\vec{k}^\prime} = (2 \pi)^3 \delta^3(\vec{k} - \vec{k}^\prime).
\end{align}
Taking derivatives of $Y$, we define vector and tensor mode functions as
\begin{align}
  Y_i &= - k^{-1} \tilde{\nabla}_i Y\,,
\\
  Y_{ij} &= k^{-2} \tilde{\nabla}_i \tilde{\nabla}_j Y + \frac{1}{3} \tilde{g}_{ij} Y.
\end{align}
Like the metric perturbation $h_{ij}$, we raise and lower indices on $Y_i$ and $Y_{ij}$ with the spatial metric only. This is useful not only for consistency with $h_{ij}$, but also because it ensures that $Y$, $Y_i$, and $Y_{ij}$ are independent of time, no matter the position of their indices.

In synchronous gauge, the metric perturbation decomposes as
\begin{align}
  h_{ij} (\vec{x}, t) = \int d^3 k \left[ \frac{h(\vec{k}, t)}{3} Y \tilde{g}_{ij} + k^2 \eta(\vec{k}, t) Y_{ij} \right].
\end{align}
Following Ma and Bertschinger \cite{ma1995}, we define a new field $\tilde{\eta}$ in momentum space by
\begin{align}
  \eta(\vec{k}, t) &= - \frac{1}{k^2} [ h(\vec{k}, t) + 6 \tilde{\eta} (\vec{k}, t)].
\end{align}
Using this definition, the metric perturbation in synchronous gauge becomes
\begin{align}
  h_{ij} = \int d^3 k \left[ \frac{h}{3} Y \tilde{g}_{ij} - (h + 6 \tilde{\eta}) Y_{ij} \right].
\end{align}
In the limit of no spatial curvature, the mode function $Y = \exp(i \vec{k} \cdot \vec{x})$ yields Ma and Bertschinger's Eq. (4).

%%%%%%%%%%%%%%%%%%%%%%%%%%%%%%%%%%%%%%
%%%%%%%%%%%%%%%%%%%%%%%%%%%%%%%%%%%%%%
%%%%%%%%%%%%%%%%%%%%%%%%%%%%%%%%%%%%%%
\section{Equations of Motion} \label{app:EOMs}

Here, we present the contribution to the $\pi$ equation of motion and the effective stress-energy tensor for a number of operators, both in synchronous gauge and Newtonian gauge. We begin by giving the equations for the background operators, then present the contribution each further operator adds to these equations.

\subsection{Background Terms}

\subsubsection{\texorpdfstring{$\pi$}{Pi} Equation of Motion}
In synchronous gauge, the equation of motion for $\pi$ from the background action [first line of Eq. \eqref{eq:piaction}] is
\begin{align}
  0 ={}&
  c \ddot{\pi}
  + c \frac{k^2}{a^2} \pi
  + \dot{\pi} \dot{c}
  + 3 H c \dot{\pi}
  + \left( \frac{\Mp}{4} \dot{\Omega} \dot{R}^{(0)} - 3 \dot{H} c \right) \pi
  + c \frac{\dot{h}}{2}
  - \frac{\Mp}{4} \dot{\Omega} \delta R. \label{eq:pimotion}
\end{align}
The perturbation to the Ricci scalar, included in the final term, is given by
\begin{align}
  \delta R = 4 H \dot{h} + \ddot{h} + \frac{4}{a^2} \tilde{\eta} \left( 3 k_0 - k^2 \right).
\end{align}
In the limit of $c(t) = 0$, the equation of motion for $\pi$ becomes algebraic in the absence of further operators, and can be solved to give
\begin{align}
  \pi = \frac{\delta R}{\dot{R}^{(0)}} \label{eq:picvanish}.
\end{align}
However, the $\left\{\pi, h, \tilde{\eta} \right\}$ system still contains one dynamical degree of freedom because of the kinetic mixing.

In Newtonian gauge, the equation of motion from the background action is given by
\begin{align}
  0 ={}&
  c \ddot{\pi}^N - c \dot{\psi}^N
  + c \frac{k^2}{a^2} \pi^N
  + \dot{c} \left(\dot{\pi}^N - \psi^N\right)
  + 3 c H \dot{\pi}^N
  - 3 c \dot{\phi}
  - 6 c H \psi^N
\nonumber\\&
  + \left( \frac{\Mp}{4} \dot{\Omega} \dot{R}^{(0)} - 3 c \dot{H} \right) \pi^N
  - \frac{\Mp}{4} \dot{\Omega} \delta R^N\,, \label{eq:piEOMN}
\end{align}
where quantities in Newtonian gauge are denoted by a superscript $N$. In Newtonian gauge, $\delta R^N$ is given by
\begin{align}
  \delta R^N = -12 (2 H^2 + \dot{H}) \psi^N - 6 H \dot{\psi}^N + 2 \frac{k^2}{a^2} \psi^N - 24 H \dot{\phi}^N - 6 \ddot{\phi}^N - 4 \frac{(k^2 - 3k_0)}{a^2} \phi^N.
\end{align}
In the appropriate limit, Eq. \eqref{eq:picvanish} still holds, using $\delta R^N$ instead of $\delta R$.

\subsubsection{Effective Stress-Energy Tensor}
The gravitational equations of motion can be obtained by varying the action \eqref{eq:basicpiaction} with respect to the metric. The resulting equations can be written in terms of the Einstein tensor, matter stress-energy tensor, and an effective dark energy stress-energy tensor as
\begin{align}
  \Mp \Omega \tensor{G}{^\mu_\nu} = \tensor{T}{^\mu_\nu^{(m)}} + \tensor{T}{^\mu_\nu^{(Q)}}. \label{eq:masterEOM}
\end{align}
We ignore the background contributions that have already been solved. The perturbed Einstein tensor is given by the following in synchronous gauge, where we use mode functions as described in Appendix \ref{app:momentum}.
\begin{align}
  \delta \tensor{G}{^0_0} &= \left[
  - H \dot{h}
  + \frac{\tilde{\eta}}{a^2} \left(2 k^2 - 6 k_0\right) \right] Y
\\
  \delta \tensor{G}{^0_i} &= \left[ 2 \dot{\tilde{\eta}} k - (\dot{h} + 6 \dot{\tilde{\eta}}) \frac{k_0}{k} \right] Y_i
\\
  \delta \tensor{G}{^i_j} &= \left[
  - \frac{\ddot{h}}{3}
  - \dot{h} H
  + \frac{2k^2}{3 a^2} \tilde{\eta}
  - \frac{2k_0}{a^2} \tilde{\eta}
  \right] \delta^i_j Y
  + \frac{1}{2} \left[
  - 3 H (\dot{h} + 6 \dot{\tilde{\eta}})
  -(\ddot{h} + 6 \ddot{\tilde{\eta}})
  + \frac{2 k^2}{a^2} \tilde{\eta}
  \right] \tensor{Y}{^i_j}
\end{align}
The scalar modes of the matter stress-energy tensor in momentum space can be decomposed in the following manner, following Kodama and Sasaki \cite{kodama1984}.
\begin{align}
  \tensor{T}{^0_0}^{(m)} &= - (\bar{\rho}_m + \delta \rho Y)
\\
  \tensor{T}{^0_i}^{(m)} &= (\bar{\rho}_m + \bar{P}_m) v Y_i
\\
  \tensor{T}{^i_j}^{(m)} &= \left(\bar{P}_m + \delta P Y \right) \delta^i_j + \bar{P}_m \Pi \tensor{Y}{^i_j}
\end{align}
Working to linear order, the components of Eq. \eqref{eq:masterEOM} are as follows, where we again use synchronous gauge and focus on terms arising from the background operators only.

\noindent \textbf{Time-time (Synchronous)}
\begin{align}
  \frac{2 \Mp}{a^2} \Omega \left[
    k^2 \tilde{\eta}
  - \frac{a^2 H}{2} \dot{h}
  - 3 k_0 \tilde{\eta} \right]
  ={}& - \delta \rho
  - \dot{\rho}_Q \pi
  - 2 c \dot{\pi}
\nonumber\\&
  + \Mp \dot{\Omega} \left[
    3 \pi \left(H^2 - \dot{H} + \frac{k_0}{a^2}\right)
  + 3 H \dot{\pi}
  + \frac{k^2}{a^2} \pi
  + \frac{\dot{h}}{2}
  \right] \label{eq:timetime}
\end{align}
\textbf{Time-space (Synchronous)}
\begin{align}
  \Mp \Omega \left[ 2 k^2 \dot{\tilde{\eta}} - k_0 (\dot{h} + 6 \dot{\tilde{\eta}}) \right] =
  (\bar{\rho}_m + \bar{P}_m) k v
  + \Mp \dot{\Omega} k^2 \dot{\pi}
  + \left[\rho_Q + P_Q\right] k^2 \pi
\end{align}
\textbf{Space-space trace (Synchronous)}
\begin{align}
  \Mp \Omega \left[
  - \frac{\ddot{h}}{3}
  - \dot{h} H
  + \frac{2k^2}{3 a^2} \tilde{\eta}
  - \frac{2k_0}{a^2} \tilde{\eta}
  \right] = {}& \delta P
  + \dot{P}_Q \pi
  + \left(\rho_Q + P_Q \right) \dot{\pi}
\label{eq:SStrace} \\&
  + \Mp \dot{\Omega} \left[
    \frac{\ddot{\Omega}}{\dot{\Omega}} \dot{\pi}
  + \ddot{\pi}
  + 3 H \dot{\pi}
  + \frac{2}{3} \frac{k^2}{a^2} \pi
  + \frac{\dot{h}}{3}
  + \pi \left( 3 H^2 + \frac{k_0}{a^2} \right)
  \right] \nonumber
\end{align}
\textbf{Space-space traceless (Synchronous)}
\begin{align}
  \frac{\Mp \Omega}{2} \left[
    \frac{2 k^2}{a^2} \tilde{\eta}
  - 3 H (\dot{h} + 6 \dot{\tilde{\eta}})
  -(\ddot{h} + 6 \ddot{\tilde{\eta}})
  \right]
  = \bar{P}_m \Pi
  + \Mp \dot{\Omega} \left[
    \frac{k^2}{a^2} \pi
  + \frac{1}{2} \left(\dot{h} + 6 \dot{\tilde{\eta}}\right)
  \right] \label{eq:SStraceless}
\end{align}

We can translate these into Newtonian gauge. The required gauge transformations are detailed in Appendix \ref{app:Synchronous}.

\noindent\textbf{Time-time (Newtonian)}
\begin{align}
  &\Mp \Omega(t) \left[
    \frac{2k^2 - 6k_0}{a^2} \phi^N
  + 6 H (\dot{\phi}^N + H \psi^N)
  \right]
\nonumber\\
  ={}& - \delta \rho^N
  - \dot{\rho}_Q \pi^N
  - 2 c(t) (\dot{\pi}^N - \psi^N)
\nonumber\\&
  + \Mp \dot{\Omega} \left[
  3 \pi^N \left(H^2 - \dot{H} + \frac{k_0}{a^2}\right)
  + 3 H \left(\dot{\pi}^N - \psi^N\right)
  + \frac{k^2}{a^2} \pi^N
  - 3 \left(\dot{\phi}^N + H \psi^N \right)
  \right] \label{eq:timetimeN}
\end{align}
\textbf{Time-space (Newtonian)}
\begin{align}
  \Mp \Omega(t) 2 k^2 \left[ \dot{\phi}^N + H \psi^N \right] ={}&
  (\bar{\rho}_m + \bar{P}_m) k v^N
  + \left[\rho_Q + P_Q\right] k^2 \pi^N
  + \Mp \dot{\Omega} k^2 \left(\dot{\pi}^N - \psi^N\right) \label{eq:timespaceN}
\end{align}
\textbf{Space-space trace (Newtonian)}
\begin{align}
  & \Mp \Omega \left[
  2 \ddot{\phi}
  + 2 \left(
    3 H^2
  + 2 \dot{H}
  \right) \psi
  + 2 H (\dot{\psi} + 3 \dot{\phi})
  + \frac{2 k^2}{3 a^2} (\phi  - \psi)
  \right] \nonumber
\\ = {}& \delta P^N
  + \dot{P}_Q \pi^N
  + \left(\rho_Q + P_Q \right) (\dot{\pi}^N - \psi^N) \nonumber
\\&
  + \Mp \dot{\Omega} \left[
    \frac{\ddot{\Omega}}{\dot{\Omega}} (\dot{\pi}^N - \psi^N)
  + \ddot{\pi}^N - \dot{\psi}^N
  - 2 \dot{\phi}^N
  + 3 H \dot{\pi}^N
  - 5 H \psi^N
  + \pi^N \left( 3 H^2 + \frac{k_0}{a^2} \right)
  + \frac{2}{3} \frac{k^2}{a^2} \pi^N
  \right]
\end{align}
\textbf{Space-space traceless (Newtonian)}
\begin{align}
  \Mp \Omega(t) \frac{k^2}{a^2} \left(\phi^N - \psi^N\right)
  = \bar{P}_m \Pi
  + \Mp \dot{\Omega} \frac{k^2}{a^2} \pi^N \label{eq:spacespacetracelessN}
\end{align}

The matter terms also transform in going to Newtonian gauge.
\begin{align}
  \delta \rho^N &= \delta \rho
  + \dot{\bar{\rho}}_m (\dot{h} + 6 \dot{\tilde{\eta}}) \frac{a^2}{2 k^2}
\\
  \delta P^N &= \delta P + \dot{P}_m \frac{a^2}{2k^2} (\dot{h} + 6 \dot{\tilde{\eta}})
\\
  (\bar{\rho}_m + \bar{P}_m) k v^N &= (\bar{\rho}_m + \bar{P}_m) k v
  + \left( \bar{P}_m + \bar{\rho}_m \right) \frac{a^2}{2} (\dot{h} + 6 \dot{\tilde{\eta}})
\end{align}
The matter anisotropic shear stress is invariant.

\subsection{Further Operators}
We now investigate the terms $(\delta g^{00})^2$, $\delta g^{00} \delta \tensor{K}{^\mu_\mu}$, $(\delta \tensor{K}{^\mu_\mu})^2$, $\delta \tensor{K}{^\mu_\nu} \delta \tensor{K}{^\nu_\mu}$ and $\delta g^{00} \delta R^{(3)}$, calculating their contributions to the equations of motion.

To describe contributions to the $\pi$ equation of motion \eqref{eq:pimotion}, we write the equation of motion as
\begin{align}
  0 = c \ddot{\pi} + c \frac{k^2}{a^2} \pi + \ldots + \Delta_\pi\,,
\end{align}
and state $\Delta_\pi$ for individual operators, where the ellipsis includes terms from other operators. For the stress-energy tensor, we state the contribution to $\tensor{T}{^\mu_\nu^{(Q)}}$, for which there should be no ambiguity.

\subsubsection*{Operator: $(\delta g^{00})^2$}
This term very generically arises from writing a covariant theory in our formalism. As a simple example, any model that involves $f(X)$ type terms where $X \propto g^{\mu \nu} \nabla_\mu \phi \nabla_\nu \phi$ such as $k$-essence will include this term. After implementing the St\"uckelberg trick, the action becomes the following.
\begin{align}
  S_{(\delta g^{00})^2} = \int d^4x \sqrt{-g} \left[ \frac{M_2^4(t)}{2} (\delta g^{00} - 2 \dot{\pi})^2 \right]
\end{align}
The $\pi$ equation of motion receives the following contribution in synchronous gauge.
\begin{align}
  \Delta_\pi = 2 M_2^4 \ddot{\pi} + (8 M_2^3 \dot{M}_2 + 6 H M_2^4) \dot{\pi}
\end{align}
In Newtonian gauge, this is
\begin{align}
  \Delta_\pi = 2 M_2^4 (\ddot{\pi}^N - \dot{\psi}^N)
  + (8 M_2^3 \dot{M}_2 + 6 H M_2^4) (\dot{\pi}^N - \psi^N).
\end{align}
The contribution to the stress-energy tensor is
\begin{align}
  \tensor{T}{^\mu_\nu^{(Q)}_{(\delta g^{00})^2}} = - 4 M_2^4 \dot{\pi} \delta^\mu_0 \delta^0_\nu Y = - 4 M_2^4 (\dot{\pi}^N - \psi^N) \delta^\mu_0 \delta^0_\nu Y
\end{align}
in synchronous gauge and Newtonian gauge respectively.

\subsubsection*{Operator: $\delta g^{00} \delta \tensor{K}{^\mu_\mu}$}
This term also often arises from writing covariant theories in this
formalism. For example,
Galileon terms of the form $(\nabla \phi)^2 \Box \phi$
\cite{Chow:2009fm} and more generally kinetic braiding terms
of the form $G(\phi,(\nabla \phi)^2) \Box \phi$ \cite{Deffayet:2010qz}
will generate it, while the full
expansion of quadratic curvature invariants in this formalism will
also typically generate it.

The action with the $\pi$ field is
\begin{align}
  S_{\delta g^{00} \delta K} = \int d^4x \sqrt{-g} \left[ - \frac{\bar{M}_1^3(t)}{2} (\delta g^{00} - 2 \dot{\pi}) \left(\tensor{K}{^\mu_\mu} + 3H + 3 \dot{H} \pi + \frac{\tilde{\nabla}^2 \pi}{a^2}\right) \right].
\end{align}

$\pi$ equation of motion in synchronous gauge:
\begin{align}
\Delta_\pi =
  \frac{1}{2} [(3H + \partial_t) \bar{M}_1^3] \left(- \frac{\dot{h}}{2} + 3 \dot{H} \pi - \frac{k^2 \pi}{a^2}\right)
  + \frac{\bar{M}_1^3}{2} \left(- \frac{\ddot{h}}{2} + 3 \ddot{H} \pi
  + 2 H \frac{k^2 \pi}{a^2}
  \right)
\end{align}
In Newtonian gauge:
\begin{align}
  \Delta_\pi ={}&
  \frac{1}{2} [(3H + \partial_t) \bar{M}_1^3] \left(3 \left(\dot{\phi}^N + H \psi^N\right) + 3 \dot{H} \pi^N - \frac{k^2}{a^2} \pi^N\right)
\nonumber\\&
  + \frac{\bar{M}_1^3}{2} \left(3 \ddot{\phi}^N + 6 \dot{H} \psi^N + 3 H \dot{\psi}^N
  + 3 \ddot{H} \pi^N
  + 2 H \frac{k^2}{a^2} \pi^N
  - \frac{k^2}{a^2} \psi^N
  \right)
\end{align}

Stress-energy tensor (synchronous gauge):
\begin{align}
  \tensor{T}{^0_0^{(Q)}_{\delta g^{00} \delta K}}
  &= \bar{M}_1^3 \left( \frac{\dot{h}}{2} - 3 \dot{H} \pi + \frac{k^2}{a^2} \pi + 3H \dot{\pi}
  \right) Y
\\
  \tensor{T}{^0_i^{(Q)}_{\delta g^{00} \delta K}}
  &=
  k \bar{M}_1^3 \dot{\pi} Y_i
\\
  \tensor{T}{^i_j^{(Q)}_{\delta g^{00} \delta K}}
  &=
  (3H + \partial_t) \left[\bar{M}_1^3 \dot{\pi} \right] \delta^i_j Y
\end{align}
Note that we find the $ij$ component is a factor of two different to Eq. (C.13) in Creminelli \textit{et al} \cite{Creminelli2009}. In Newtonian gauge:
\begin{align}
  \tensor{T}{^0_0^{(Q)}_{\delta g^{00} \delta K}}
  &=
  \bar{M}_1^3 \left( - 3 \dot{\phi}^N - 6H \psi^N - 3 \dot{H} \pi^N + \frac{k^2}{a^2} \pi^N + 3H \dot{\pi}^N
  \right) Y
\\
  \tensor{T}{^0_i^{(Q)}_{\delta g^{00} \delta K}}
  &=
  k \bar{M}_1^3 \left(\dot{\pi}^N - \psi^N \right) Y_i
\\
  \tensor{T}{^i_j^{(Q)}_{\delta g^{00} \delta K}}
  &=
  (3H + \partial_t) \left[\bar{M}_1^3 \left(\dot{\pi}^N - \psi^N \right) \right] \delta^i_j Y
\end{align}

\subsubsection*{Operator: $(\delta K)^2$}

The effect of this term on the behavior of the $\pi$ field has been investigated in detail in the ghost condensate limit, in particular in Ref. \cite{Cheung2008}.

Action with $\pi$ field:
\begin{align}
  S_{(\delta K)^2} = \int d^4x \sqrt{-g} \left[ - \frac{\bar{M}_2^2(t)}{2} \left(\tensor{K}{^\mu_\mu} + 3H + 3 \dot{H} \pi + \frac{\tilde{\nabla}^2 \pi}{a^2}\right)^2 \right]
\end{align}

$\pi$ equation of motion in synchronous gauge.
\begin{align}
  \Delta_\pi = \frac{\bar{M}_2^2}{2} \left(3 \dot{H} - \frac{k^2}{a^2}\right) \left(- \frac{\dot{h}}{2} + 3 \dot{H} \pi - \frac{k^2}{a^2} \pi \right)
\end{align}
In Newtonian gauge:
\begin{align}
  \Delta_\pi = \frac{\bar{M}_2^2}{2} \left(3 \dot{H} - \frac{k^2}{a^2}\right) \left(3 \left(\dot{\phi}^N + H \psi^N\right) + 3 \dot{H} \pi^N - \frac{k^2}{a^2} \pi^N\right)
\end{align}

Stress-energy tensor in synchronous gauge:
\begin{align}
  \tensor{T}{^0_0^{(Q)}}_{(\delta K)^2}
  &= 3H \bar{M}_2^2 \left( \frac{\dot{h}}{2} - 3 \dot{H} \pi + \frac{k^2}{a^2} \pi \right) Y
\\
  \tensor{T}{^0_i^{(Q)}}_{(\delta K)^2}
  &= k \bar{M}_2^2 \left(\frac{\dot{h}}{2} - 3 \dot{H} \pi + \frac{k^2}{a^2} \pi \right) Y_i
\\
  \tensor{T}{^i_j^{(Q)}}_{(\delta K)^2}
  &= \left(3H + \partial_t \right)
  \left[\bar{M}_2^2 \left(\frac{\dot{h}}{2} - 3 \dot{H} \pi + \frac{k^2}{a^2} \pi \right)\right] \delta ^i_j Y
\end{align}
Again, we find the $ij$ component to be a factor of two different from Creminelli's Eq. (C.13). Newtonian gauge:
\begin{align}
  \tensor{T}{^0_0^{(Q)}}_{(\delta K)^2}
  &= 3H \bar{M}_2^2 \left( - 3 \dot{\phi}^N - 3 H \psi^N - 3 \dot{H} \pi^N + \frac{k^2}{a^2} \pi^N \right) Y
\\
  \tensor{T}{^0_i^{(Q)}}_{(\delta K)^2}
  &= k \bar{M}_2^2 \left(- 3 \dot{\phi}^N - 3 H \psi^N - 3 \dot{H} \pi^N + \frac{k^2}{a^2} \pi^N \right) Y_i
\\
  \tensor{T}{^i_j^{(Q)}}_{(\delta K)^2}
  &= \left(3H + \partial_t \right)
  \left[\bar{M}_2^2 \left(- 3 \dot{\phi}^N - 3 H \psi^N - 3 \dot{H} \pi^N + \frac{k^2}{a^2} \pi^N \right) \right] \delta ^i_j Y
\end{align}

\subsubsection*{Operator: $\delta \tensor{K}{^\mu_\nu} \delta \tensor{K}{^\nu_\mu}$}
In terms of the effect on the behavior of the $\pi$ field, this term is qualitatively similar to $(\delta K)^2$. The gravitational behavior is somewhat more complicated, however. Note that this term is projected to spatial indices only for the purposes of introducing the $\pi$ field.
\begin{align}
  S_{\delta \tensor{K}{^\mu_\nu} \delta \tensor{K}{^\nu_\mu}} = \int d^4x \sqrt{-g} {}& \bigg\{
    - \frac{\bar{M}_3^2 (t)}{2}
  \left(\delta \tensor{K}{^i_j} + \dot{H} \pi \delta^i_j
  + \frac{\tilde{g}^{i k}}{a^2} \tilde{\nabla}_k \tilde{\nabla}_j \pi \right)
  \left(\delta \tensor{K}{^j_i} + \dot{H} \pi \delta^j_i
  + \frac{\tilde{g}^{j l}}{a^2} \tilde{\nabla}_l \tilde{\nabla}_i \pi \right)
\bigg\}
\end{align}

The $\pi$ equation of motion in synchronous gauge is
\begin{align}
\Delta_\pi = \bar M_3^2\Big[
\Big(\frac{k^4}{2a^4} - \frac{k^2}{a^2} \dot{H} + \frac{3}{2} \dot{H}^2 - \frac{k^2 k_0}{a^4} \Big) \pi
+ \Big(\frac{k^2}{4a^2}-\frac{\dot H}{4} - \frac{k_0}{2 a^2}\Big)\dot h
+ \frac{k^2 - 3 k_0}{a^2} \dot{\tilde\eta}\Big] \,,
\end{align}
and in Newtonian gauge,
\begin{align}
\Delta_\pi = \bar M_3^2\Big[
\Big(\frac{k^4}{2a^4} - \frac{k^2}{a^2} \dot{H} + \frac{3}{2} \dot{H}^2 - \frac{k^2 k_0}{a^4} \Big) \pi^N
+ \Big(\frac{3 H \dot{H}}{2} - \frac{H k^2}{2 a^2}\Big) \psi^N
+ \Big(\frac{3 \dot{H}}{2} - \frac{k^2}{2 a^2} \Big) \dot{\phi}^N
\Big] \,.
\end{align}

Obtaining the stress-energy tensor is a lengthy calculation. In terms of the quantity
\begin{align}
  \tensor{F}{^j_i} = - \bar{M}_3^2 (t)
    \left( \delta \tensor{K}{^j_i}
    + \dot{H} \pi \delta^j_i
    + \frac{\tilde{g}^{j k}}{a^2} \tilde{\nabla}_k \tilde{\nabla}_i \pi
    \right)
\end{align}
with spatial indices only, the components of the stress-energy tensor (in arbitrary gauge) are
\begin{align}
  \tensor{T}{^0_0^{(Q)}}_{\delta \tensor{K}{^\mu_\nu} \delta \tensor{K}{^\nu_\mu}}
  ={}& H \tensor{F}{^i_i}
\\
  \tensor{T}{^0_i^{(Q)}}_{\delta \tensor{K}{^\mu_\nu} \delta \tensor{K}{^\nu_\mu}}
  ={}& - \tilde{\nabla}_j \tensor{F}{^j_i}
\\
  \tensor{T}{^i_j^{(Q)}}_{\delta \tensor{K}{^\mu_\nu} \delta \tensor{K}{^\nu_\mu}}
  ={}& (3H + \partial_t) \tensor{F}{^i_j} \,.
\end{align}
In synchronous gauge, this is
\begin{align}
  \tensor{T}{^0_0^{(Q)}}_{\delta \tensor{K}{^\mu_\nu} \delta \tensor{K}{^\nu_\mu}}
  ={}& H \bar{M}_3^2 (t)
    \left(
      \frac{1}{2} \dot{h}
    - 3 \dot{H} \pi
    + \frac{k^2}{a^2} \pi
    \right) Y
\\
  \tensor{T}{^0_i^{(Q)}}_{\delta \tensor{K}{^\mu_\nu} \delta \tensor{K}{^\nu_\mu}}
  ={}& \bar{M}_3^2 (t)
    \left(
    - \frac{1}{6}\dot{h} - \left(\frac{1}{3} - \frac{k_0}{k^2}\right) (\dot{h} + 6 \dot{\tilde{\eta}})
    + \dot{H} \pi
    + \frac{2 k_0 - k^2}{a^2} \pi
    \right) (- k Y_i)
\\
  \tensor{T}{^i_j^{(Q)}}_{\delta \tensor{K}{^\mu_\nu} \delta \tensor{K}{^\nu_\mu}}
  ={}& - \bar{M}_3^2 (t) \left(3 H + 2 \frac{\dot{\bar{M}}_3}{\bar{M}_3} + \partial_t\right) \times
\nonumber\\
& \qquad \left( - \frac{1}{6} \delta^i_j Y \dot{h} + \tensor{Y}{^i_j} \frac{1}{2} (\dot{h} + 6 \dot{\tilde{\eta}})
    + \dot{H} \pi \delta^i_j
    + \frac{k^2}{a^2} \left( \tensor{Y}{^i_j} - \frac{1}{3} \delta^i_j Y \right) \pi
    \right) \, .
\end{align}
In Newtonian gauge, this becomes
\begin{align}
  \tensor{T}{^0_0^{(Q)}}_{\delta \tensor{K}{^\mu_\nu} \delta \tensor{K}{^\nu_\mu}}
  ={}& H \bar{M}_3^2 (t)
    \left( - 3 H \psi^N - 3 \dot{\phi}^N
    - 3 \dot{H} \pi^N
    + \frac{k^2}{a^2} \pi^N
    \right) Y
\\
  \tensor{T}{^0_i^{(Q)}}_{\delta \tensor{K}{^\mu_\nu} \delta \tensor{K}{^\nu_\mu}}
  ={}& \bar{M}_3^2 (t)
    \left( H \psi^N + \dot{\phi}^N
    + \dot{H} \pi^N
    + \frac{2 k_0 - k^2}{a^2} \pi^N
    \right) (- k Y_i)
\\
  \tensor{T}{^i_j^{(Q)}}_{\delta \tensor{K}{^\mu_\nu} \delta \tensor{K}{^\nu_\mu}}
  ={}& - \bar{M}_3^2 (t) \left(3 H + 2 \frac{\dot{\bar{M}}_3}{\bar{M}_3} + \partial_t\right) \times
\nonumber\\
& \quad    \left( (H \psi^N + \dot{\phi}^N) \delta^i_j Y
    + \dot{H} \pi^N Y \delta^i_j
    + \frac{k^2}{a^2} \left( \tensor{Y}{^i_j} - \frac{1}{3} \delta^i_j Y \right) \pi^N
    \right) \,.
\end{align}

\subsubsection*{Operator: $\delta g^{00} \delta R^{(3)}$}
This term is required to fully describe perturbations in Horndeski's theory. Note that $R^{(3)}$ is the Ricci scalar of the spatial metric $g_{ij}$, including the scale factor.

The action with the $\pi$ field introduced is
\begin{align}
  S_{\delta g^{00} \delta R^{(3)}} = \int d^4x \sqrt{-g} {}& \bigg[
    \frac{\hat{M}^2 (t)}{2} \left(\delta g^{00} - 2 \dot{\pi} \right) \left(\delta R^{(3)} + 4 H \frac{\tilde{\nabla}^2 \pi}{a^2} + 12 H \frac{k_0}{a^2} \pi \right)
  \bigg].
\end{align}
The perturbed three-dimensional Ricci scalar is
\begin{align}
  \delta R^{(3)} &= 4 \frac{(3 k_0 - k^2)}{a^2} \tilde{\eta} = 4 \frac{(3 k_0 - k^2)}{a^2} \phi^N
\end{align}
in synchronous and Newtonian gauge, respectively.

The $\pi$ equation of motion is as follows, in synchronous and Newtonian gauge.
\begin{align}
  \Delta_\pi = {}& 2 \frac{(k^2 - 3 k_0)}{a^2} \left[\left( H \hat{M}^2 + 2 \hat{M} \dot{\hat{M}} \right) \tilde{\eta}
  + \hat{M}^2 \dot{\tilde{\eta}} + \left( H^2 \hat{M}^2 + 2 H \hat{M} \dot{\hat{M}} + \hat{M}^2 \dot{H} \right) \pi \right]
\\
= &{} 2 \frac{(k^2 - 3 k_0)}{a^2} \left[\left( H \hat{M}^2 + 2 \hat{M} \dot{\hat{M}} \right) \phi^N
  + \hat{M}^2 \dot{\phi}^N + \left( H^2 \hat{M}^2 + 2 H \hat{M} \dot{\hat{M}} + \hat{M}^2 \dot{H} \right) \pi^N + H \hat{M}^2 \psi^N \right]
\end{align}

The stress-energy tensor in synchronous gauge is
\begin{align}
  \tensor{T}{^0_0^{(Q)}}_{\delta g^{00} \delta R^{(3)}} &= 4 \hat{M}^2 \frac{3 k_0 - k^2}{a^2} (\tilde{\eta} + H \pi) Y
\\
  \tensor{T}{^0_i^{(Q)}}_{\delta g^{00} \delta R^{(3)}} &= 0
\\
  \tensor{T}{^i_j^{(Q)}}_{\delta g^{00} \delta R^{(3)}} &=
  - 2 \hat{M}^2 \dot{\pi} \left[ \frac{2}{3} \frac{k^2 - 3 k_0}{a^2} Y \delta^i_j + \frac{k^2}{a^2} \tensor{Y}{^i_j} \right].
\end{align}
In Newtonian gauge, this becomes
\begin{align}
  \tensor{T}{^0_0^{(Q)}}_{\delta g^{00} \delta R^{(3)}} &= 4 \hat{M}^2 \frac{3 k_0 - k^2}{a^2} (\phi^N + H \pi^N) Y
\\
  \tensor{T}{^0_i^{(Q)}}_{\delta g^{00} \delta R^{(3)}} &= 0
\\
  \tensor{T}{^i_j^{(Q)}}_{\delta g^{00} \delta R^{(3)}} &=
  2 \hat{M}^2 (\psi^N - \dot{\pi}^N) \left[ \frac{2}{3} \frac{k^2 - 3 k_0}{a^2} Y \delta^i_j + \frac{k^2}{a^2} \tensor{Y}{^i_j} \right].
\end{align}

\subsubsection*{Operator: $(g^{\mu \nu} + n^\mu n^\nu) \partial_\mu \delta g^{00} \partial_\nu \delta g^{00}$}
Gubitosi \textit{et al.} \cite{Gubitosi2012} have pointed out that this term is the leading order term in the expansion of Ho\v{r}ava-Lifshitz gravity. With the $\pi$ field introduced, the action is
\begin{align}
  S_{(g^{\mu \nu} + n^\mu n^\nu) \partial_\mu \delta g^{00} \partial_\nu \delta g^{00}} = \int d^4x \sqrt{-g} \,
    m_2^2 (t) (g^{\mu \nu} + n^\mu n^\nu) \partial_\mu (g^{00} - 2 \dot{\pi}) \partial_\nu (g^{00} - 2 \dot{\pi}).
\end{align}

The $\pi$ equation of motion is
\begin{align}
  \Delta_\pi = 4 (3H + \partial_t) \left(m_2^2 \frac{k^2}{a^2} \dot{\pi} \right)
\end{align}
in synchronous gauge, and
\begin{align}
  \Delta_\pi = 4 (3H + \partial_t) \left(m_2^2 \frac{k^2}{a^2} (\dot{\pi}^N - \psi^N) \right)
\end{align}
in Newtonian gauge.

The stress-energy tensor is
\begin{align}
  \tensor{T}{^\mu_\nu^{(Q)}}_{(g^{\mu \nu} + n^\mu n^\nu) \partial_\mu \delta g^{00} \partial_\nu \delta g^{00}}
  = -8 m_2^2 \frac{k^2}{a^2} \dot{\pi} Y \delta_0^\mu \delta^0_\nu
  = 8 m_2^2 \frac{k^2}{a^2} (\psi^N - \dot{\pi}^N) Y \delta_0^\mu \delta^0_\nu
\end{align}
in synchronous and Newtonian gauge, respectively.

\bibliography{darkenergyeft}
\bibliographystyle{JHEP}

\end{document}